\pdfoutput=1
\newcommand*{\ATLASLATEXPATH}{}
\documentclass[cernpreprint,texlive=2016,txfonts,USenglish]{\ATLASLATEXPATH atlasdoc}

\usepackage{\ATLASLATEXPATH atlaspackage}
\usepackage{\ATLASLATEXPATH atlasbiblatex}
\usepackage{\ATLASLATEXPATH atlasphysics}
\usepackage{\ATLASLATEXPATH atlasheavyion}
\usepackage{\ATLASLATEXPATH atlasunit}
\usepackage{multirow}

\usepackage{ridgeLongPaper-defs}

\AtlasTitle{Measurements of long-range azimuthal anisotropies and
associated Fourier coefficients for $pp$ collisions at $\sqrt{s}=5.02$ 
and 13~TeV and $p$+Pb  collisions at $\sqrt{s_{\mathrm{NN}}}=5.02$~TeV
with the ATLAS detector} 

\PreprintIdNumber{CERN-EP-2016-200}

\AtlasJournalRef{Phys. Rev. C 96 (2017) 024908}
\AtlasDOI{10.1103/PhysRevC.96.024908}

\AtlasAbstract{%
ATLAS measurements of two-particle
correlations are presented for $\sqrt{s} = 5.02$ and 13~TeV $pp$ collisions and for
$\sqrt{s_{\mathrm{NN}}} = 5.02$~TeV $p$+Pb collisions at the LHC. The correlation
functions are measured as a function of relative azimuthal angle
$\Delta \phi$, and pseudorapidity separation $\Delta \eta$, using charged
particles detected within the pseudorapidity interval
$|\eta|{<}2.5$. Azimuthal modulation in the long-range component of
the correlation function, with $|\Delta\eta|{>}2$, is studied using a template
fitting procedure to remove a ``back-to-back'' contribution to the
correlation function that primarily arises from hard-scattering
processes. In addition to the elliptic, $\cos{(2\Delta\phi)}$, modulation
observed in a previous measurement, the $pp$ correlation functions
exhibit significant $\cos{(3\Delta\phi)}$ and $\cos{(4\Delta\phi)}$ modulation.
The Fourier coefficients $v_{n,n}$ associated with the
$\cos{(n\Delta\phi)}$ modulation of the correlation functions for $n =$
2--4 are measured as a function of charged-particle
multiplicity and charged-particle transverse momentum.
The Fourier coefficients are observed to be compatible with
$\cos{(n\phi)}$ modulation of per-event single-particle azimuthal
angle distributions. The single-particle Fourier
coefficients $v_n$ are measured as a
function of charged-particle multiplicity,
and charged-particle transverse momentum for $n {=} $ 2--4. 
The integrated luminosities used in this analysis are, 64~$\mathrm{nb^{-1}}$ for the 
  $\sqrt{s}=13$~TeV $pp$ data, 170~$\mathrm{nb^{-1}}$ for the $\sqrt{s}=5.02$~TeV $pp$ data
  and 28~$\mathrm{nb^{-1}}$ for the $\sqrt{s_{\mathrm{NN}}}=5.02$~TeV $p$+Pb data.
}

\hypersetup{pdftitle={ATLAS draft},pdfauthor={The ATLAS Collaboration}}

\newcommand{\papertype}{paper}

\begin{document}

\maketitle

\section{Introduction}
\label{sec:intro}
Observations of azimuthal anisotropies in the angular distributions of
  particles produced in proton--lead (\pPb) collisions at the
  LHC \cite{Abelev:2012ola,HION-2012-13,Chatrchyan:2013nka,HION-2013-04,Khachatryan:2015waa} 
  and in deuteron--gold (\dAu)~\cite{Adare:2013piz,Adare:2014keg,Adamczyk:2015xjc} 
  and ${}^3\mathrm{He}{+}\mathrm{Au}$~\cite{Adare:2015ctn} 
  collisions at RHIC have garnered much interest due to the remarkable 
  similarities between the phenomena observed in those colliding systems 
  and the effects of collective expansion seen in the \PbPb\ and \AuAu\ 
  collisions~\cite{ALICE:2011ab,Aamodt:2011by,Chatrchyan:2012wg,HION-2011-01,Chatrchyan:2013nka,}.\footnote{However,
    Ref.\,\cite{Adamczyk:2015xjc} argues that the observed correlations may
    be due to poorly understood hard-scattering contributions.}
The most intriguing feature of the azimuthal anisotropies is the ``ridge'': 
    an enhancement in the production of particles with small azimuthal angle ($\phi$) 
   separation which extends over a large range of pseudorapidity ($\eta$) 
   separation~\cite{Khachatryan:2010gv,CMS:2012qk,Abelev:2012ola,HION-2012-13}. 
In \PbPb~\cite{ALICE:2011ab,Aamodt:2011by,Chatrchyan:2012wg,HION-2011-01,Chatrchyan:2013nka,} 
and \pPb\ \cite{HION-2012-13,Abelev:2012ola,Chatrchyan:2013nka} collisions, the ridge is  
understood to result from sinusoidal modulation of the single-particle
azimuthal angle distributions, and the characteristics of the modulation,
for example the \pT\ dependence~\cite{Kozlov:2014fqa,},
are remarkably similar in the two systems~\cite{HION-2013-04}. 

While the modulation of the azimuthal angle distributions in
\PbPb\ collisions is understood to result from the geometry of the initial
state and the imprinting of that geometry on the angular distributions
of the particles by the collective expansion (see
e.g.~\cite{Huovinen:2001cy,Romatschke:2003ms,Heinz:2013th} and 
references therein), there is, as yet, no
consensus that the modulation observed in \pPb\ collisions results
from the same mechanism. Indeed, an alternative explanation for the
modulation using perturbative QCD and assuming saturated parton
distributions in the lead nucleus is capable of reproducing many
features of the \pPb\ data~\cite{dumitru:2010iy,Ryskin:2011qh,
Dusling:2012iga,Tribedy:2011aa,Dusling:2012wy,Dusling:2013oia,Dumitru:2014vka,
Noronha:2014vva,Dumitru:2015cfa,Lappi:2015vha}. Nonetheless, because of the many
similarities between the \pPb\ and \PbPb\ observations, 
extensive theoretical and experimental effort has been devoted to
address the question of whether the strong-coupling physics 
understood to be responsible for the collective dynamics in
\NucNuc\ collisions may persist in smaller systems~\cite{Bozek:2011if,Bozek:2012gr,Bzdak:2013zma,Bozek:2013uha,Shuryak:2013ke,Bozek:2013ska,Qin:2013bha,Basar:2013hea,Habich:2015rtj,Chesler:2015bba,Chesler:2016ceu,}.

A recent study by the ATLAS Collaboration of two-particle angular correlations in 
proton--proton~(\pp) collisions at center-of-mass energies of
$\sqs= 13$ and 2.76~\TeV\
obtained results that are consistent with
the presence of an elliptic or $\cos{(2\phi)}$ modulation of the per-event single
particle azimuthal angle  distributions~\cite{HION-2015-09}. 
This result suggests that
the ridge previously observed in $\sqs = 7$~\TeV\ \pp\ collisions~\cite{
Khachatryan:2010gv}
results from modulation of the single-particle azimuthal angle
distributions similar to that seen in \PbPb\ and
\pPb\ collisions. Indeed, the \pT\ dependence of the
modulation was similar to that observed in the other systems.
Unexpectedly, the amplitude of the modulation relative to the
average differential particle yield $\langle dN/d\phi \rangle$, was observed to be
constant, within uncertainties, as a function of the charged particle
multiplicity of the \pp\ events and to be consistent between the two
energies, suggesting that the modulation is an intrinsic feature of
high-energy \pp\ collisions. These results provide further urgency to
address the question of whether strong coupling and collective
dynamics play a significant role in small systems, including the
smallest systems accessible at collider energies --
\pp\ collisions. Since the elliptic modulation observed in the
\pp\ data is qualitatively similar to that seen in \pPb\ collisions, a
direct, quantitative comparison of \pp\ and \pPb\ measurements is
necessary for evaluating whether the phenomena are related.

The modulation of the single-particle azimuthal angle distributions in \NucNuc, 
$p/d$+A, and, most recently, \pp\ collisions is usually
characterized using a set of Fourier coefficients $v_n$, that
describe the relative amplitudes of the sinusoidal components of
the single-particle distributions. More explicitly, the azimuthal
angle distributions of the particles are parameterized according to: 

\begin{equation}
\frac{dN}{d\phi} = \left\langle \frac{dN}{d\phi} \right\rangle \left(1 +  \sum_n
2v_n \cos{\left[n\left(\phi - \Psi_n\right)\right]}\right),
\label{eq:vnsingle}
\end{equation}

where the average in the equation indicates an average over azimuthal
angle. Here, $\Psi_n$ represents one of the $n$ angles at which the
$n$th-order harmonic is maximum; it is frequently referred to as the
event-plane angle for the $n$th harmonic. In \PbPb\ collisions, $n
= 2$ modulation is understood to primarily result from an elliptic
anisotropy of the initial state for collisions with non-zero impact
parameter; that anisotropy is subsequently imprinted onto the angular
distributions of the produced particles by the collective evolution of
the medium, producing an elliptic modulation of the produced particle
azimuthal angle distributions in each
event~\cite{Sorge:1996pc,Huovinen:2001cy,Kolb:2001qz}. The higher ($n
> 2$)  harmonics are understood to result 
from position-dependent fluctuations in the initial-state energy
density which produce higher-order spatial eccentricities that
similarly get converted into sinusoidal
modulation of the single-particle $dN/d\phi$ distribution 
by the collective dynamics~\cite{Alver:2010gr,Schenke:2010rr,Holopainen:2010gz,Alver:2010dn,Petersen:2010cw,Qin:2010pf,Gardim:2011xv,Schenke:2012wb,}. 
Significant
$v_n$ values have been observed in \PbPb\ (\pPb) collisions up to $n =
6$ \cite{HION-2011-01} ($n = 5$ \cite{HION-2013-04}). An important, outstanding question is whether $n > 2$
modulation is present in \pp\ collisions. 

The \vnn coefficients can be measured using two-particle
angular correlation functions, which, when evaluated as a function of
$\dphi \equiv \phi^a - \phi^b$, where $a$ and $b$ represent the two
particles used to construct the correlation function, have an
expansion similar to that in Eq.\,\eqref{eq:vnsingle}:

\begin{equation}
\frac{dN_{\mathrm{pair}}}{d\dphi} = \left\langle \frac{dN_{\mathrm{pair}}}{d\dphi} \right\rangle \left[1 +  \sum_n
2v_{n,n} \cos{\left(n\dphi\right)}\right].
\label{eq:vnpair}
\end{equation}

If the modulation of the two-particle correlation function arises
solely from the modulation of the single-particle distributions,
then, $v_{n,n} = v_n^2$.  Often, the two-particle
correlations are measured using different transverse momentum (\pT) ranges
for particles $a$ and $b$. Since the modulation is observed to vary
with \pT, then 
\begin{equation}
v_{n,n}(\pta,\ptb) = v_n (\pta) v_n (\ptb) 
\label{eq:factorone}
\end{equation}
if the
modulation of the correlation function results solely from
single-particle modulation.\footnote{See Refs.\,\cite{Gardim:2012im,Khachatryan:2015oea} for analyses of the breakdown of factorization.} This ``factorization'' hypothesis can be
tested experimentally by measuring $v_{n,n}(\pta, \ptb)$ for different
ranges of $\ptb$ and estimating $v_n(\pta)$ using

\begin{equation}
v_n(\pta) = v_{n,n}(\pta, \ptb) / \sqrt{v_{n,n}(\ptb, \ptb)} 
\label{eq:factortwo}
\end{equation}

and evaluating whether $v_n(\pta)$ depends on the choice of \ptb. 

In addition to the sinusoidal modulation, the
two-particle correlation functions include contributions from
hard-scattering processes that produce a jet peak centered at $\dphi =
\deta = 0$ and a dijet enhancement at $\dphi = \pi$ that extends over a
wide range of \deta. The jet peak can be avoided by studying the
long-range part of the correlation function, which is typically
chosen to be $|\deta| > 2$.  Because the dijet contribution to the
two-particle correlation function is not localized in \deta, that
contribution has to be subtracted from the measured correlation
function, typically using the correlation function measured in
low-multiplicity (``peripheral'') events. Different peripheral
subtraction methods have been applied for the \pPb\ measurements in
the literature~\cite{HION-2012-13,HION-2013-04}; all of them relied on the ``zero yield at
minimum'' (ZYAM)~\cite{HION-2012-13,HION-2013-04} hypothesis to subtract an assumed flat
combinatoric component from the peripheral reference correlation
function. These methods were found to be
inadequate for \pp\ collisions, where the amplitude of the dijet
enhancement at $\dphi = \pi$ is much larger than the (absolute)
amplitude of the sinusoidal modulation. For the measurements in
Ref.\,\cite{HION-2015-09}, a template fitting method, described below, was developed
which is better suited for extracting a small sinusoidal modulation
from the data. Application of the template fitting method to the
\pp\ data provided an excellent description of the measured
correlation functions. It also indicated substantial bias resulting
from the application of the ZYAM-subtraction procedure to the peripheral
reference correlation function due to the non-zero $v_{2,2}$ 
in low-multiplicity events. As a result, the measurements presented in
Ref.\,\cite{HION-2015-09} were obtained without using ZYAM subtraction. However, the
previously published \pPb\ data \cite{HION-2013-04} may be susceptible to an unknown bias
due to the use of the ZYAM method. Thus, a reanalysis of the
\pPb\ data is both warranted and helpful in making comparisons between
\pp\ and \pPb\ data.

To address the points raised above, this \papertype\ extends previous measurements
  of two-particle correlations in \pp\ collisions at $\sqs = 13$~\TeV\
  using additional data acquired by
  ATLAS subsequent to the measurements in Ref.\,\cite{HION-2015-09} 
  and provides new measurements of such correlations
  in \pp\ collisions at $\sqs = 5.02$~\TeV. It also presents a reanalysis of
  two-particle correlations in 5.02~\TeV\ \pPb\ collisions and presents
  a direct comparison between the \pp\ and \pPb\ data at the same
  per-nucleon center-of-mass energy as well as a comparison between the
  \pp\ data at the two energies.
Two-particle Fourier coefficients $v_{n,n}$ are
measured, where statistical precision allows, for $n = 2, 3,$ and 4 as a
function of charged-particle multiplicity and transverse
energy. Measurements are performed for different  \pta\ and
\ptb\ intervals and the factorization of the resulting $v_{n,n}$
values is tested. 

This \papertype\ is organized as follows. 
Section\,\ref{sec:detector} gives a brief overview of the ATLAS detector 
   subsystems and triggers used in this analysis.
Section\,\ref{sec:data} describes the data sets, and the offline 
   selection criteria used to select events and reconstruct
   charged-particle tracks.
The variables used to characterize the ``event activity'' of the \pp and \pPb
   collisions are also described.
Section\,\ref{sec:corr} gives details of the 
   two-particle correlation method.
Section\,\ref{sec:fits} describes the template fitting of 
   the two-particle correlations, which was originally 
   developed in Ref.\,\cite{HION-2015-09}.
The template fits are used to extract the Fourier harmonics \vnn 
  (Eq.\,\eqref{eq:vnpair}) of the long-range correlation, and the 
  factorization of the \vnn into single-particle harmonics \vn 
  (Eq.\,\eqref{eq:factorone}) is studied.
The stability of the \vnn as a function of the pseudorapidity separation between
  the charged-particle pairs is also checked.
Section\,\ref{sec:systematics} describes the 
  systematic uncertainties associated with the measured \vnn.
Section\,\ref{sec:results} presents the main results of the analysis, which are the 
  \pT and event-activity dependence of the single-particle harmonics, $v_n$.
Detailed comparisons of the \vn between the three data sets: 13~\TeV\ \pp, 
  5.02~\TeV\ \pp, and 5.02~\TeV\ \pPb are also shown. 
Section\,\ref{sec:conclusion} gives a summary of the main results and observations.

\section{Experiment}
\label{sec:detector}
\subsection{ATLAS detector}
The measurements presented in this \papertype\ were performed 
  using the ATLAS~\cite{PERF-2007-01} inner detector~(ID), minimum-bias trigger
  scintillators~(MBTS), calorimeter, zero-degree 
  calorimeters~(ZDC), and the trigger and data acquisition
  systems. 
The ID detects charged particles within the pseudorapidity range\footnote{ATLAS 
    uses a right-handed coordinate system with its origin at the 
    nominal interaction point (IP) in the center of the detector 
    and the $z$-axis along the beam pipe. The $x$-axis points from 
    the IP to the center of the LHC ring, and the $y$-axis points upward. 
    Cylindrical coordinates $(r,\phi)$ are used in the transverse plane,
    $\phi$ being the azimuthal angle around the $z$-axis. 
    The pseudorapidity is defined in terms of the polar angle $\theta$ 
    as $\eta=-\ln\tan(\theta/2)$.} 
  $|\eta|$<2.5 using a combination of silicon pixel detectors including
  the ``insertable B-layer''~(IBL)~\cite{Capeans:1291633,ATLAS:1451888} 
  that was installed between Run\,1 (2009--2013) and Run\,2, silicon microstrip 
  detectors~(SCT), and  a straw-tube transition radiation tracker~(TRT),
  all immersed in a 2\,T axial magnetic field~\cite{IDET-2010-01}. 
The MBTS system detects charged particles over $2.07 < |\eta| < 3.86$
   using two hodoscopes on each side of the detector, positioned at $z = \pm 3.6~{\mathrm{m}}$. 
These hodoscopes were rebuilt between Run\,1 and Run\,2.
The ATLAS calorimeter system consists of a liquid argon (LAr) electromagnetic (EM)  
   calorimeter covering $|\eta|<3.2$, a steel--scintillator sampling hadronic
   calorimeter covering $|\eta|<1.7$, a LAr hadronic calorimeter covering 
   $1.5<|\eta|<3.2$, and two LAr electromagnetic and hadronic forward calorimeters 
   (FCal) covering $3.2<|\eta|<4.9$. 
The ZDCs, situated ${\approx}\pm140$~m from the nominal IP, detect neutral  
  particles, mostly neutrons and photons, with $|\eta|$>8.3.
The ZDCs use tungsten plates as absorbers, and quartz rods sandwiched
  between the tungsten plates as the active medium.

\subsection{Trigger}

The ATLAS trigger system~\cite{PERF-2011-02} consists of a Level-1 (L1)
   trigger implemented using a combination of dedicated electronics and
   programmable logic, and a software-based high-level trigger (HLT).
Due to the large interaction rates, only a small fraction of
   minimum-bias events could be recorded for all three data sets. 
The configuration of the minimum-bias (MB) triggers varied between
    the different data sets. 
Minimum-bias \pPb\ events were selected by requiring a hit 
   in at least one MBTS counter on each side (MBTS\_1\_1) or a signal in the ZDC on the 
   Pb-fragmentation side with the trigger threshold set just below
   the peak corresponding to a single neutron.
In the 13~\TeV\ \pp data, MB events were selected by a L1 
   trigger that requires a signal in at least one MBTS counter (MBTS\_1). 
In the 5.02~\TeV\ \pp data, MB events were selected using the logical OR of 
   the MBTS\_1, MBTS\_1\_1,
   and a third trigger that required at least one reconstructed track at the HLT.
In order to increase the number of events having high charged-particle 
   multiplicity, several high-multiplicity (\HMT) triggers were implemented. 
These apply a L1 requirement on either the transverse energy (\et) 
   in the calorimeters or on the number of hits in the MBTS, and an HLT 
   requirement on the multiplicity of HLT-reconstructed charged-particle tracks.  
That multiplicity, \nchrecHLT, is evaluated for tracks having
   $\pT$>0.4~\GeV\ that are associated with the reconstructed vertex with 
   the highest multiplicity in the event. 
This last requirement suppresses the selection of events with multiple collisions (pileup), 
   as long as the collision vertices are not so close as to be indistinguishable. 
The HMT trigger configurations used in this analysis are summarized in Table~\ref{tab:pPb_HMT_triggers}.

\begin{table}
\begin{centering}
\begin{tabular}{|cc|cc|cc|}
\hline
\multicolumn{2}{|c|}{\pp 13~\TeV}     & \multicolumn{2}{c|}{\pp 5.02~\TeV}      &\multicolumn{2}{c|}{\pPb}            \tabularnewline\hline
 L1             &          HLT       & L1             &          HLT       &L1              &         HLT        \tabularnewline\hline
\rule{0pt}{2.5ex} MBTS           &\nchrecHLT$\geq$60  &\lOnecal>5~\GeV\  &\nchrecHLT$\geq$60  &\lOnefcal>10~\GeV&\nchrecHLT$\geq$100 \tabularnewline
\rule{0pt}{2.5ex}\lOnecal>10~\GeV\ &\nchrecHLT$\geq$90  &\lOnecal>10~\GeV\ & \nchrecHLT$\geq$90 &\lOnefcal>10~\GeV&\nchrecHLT$\geq$130 \tabularnewline
\rule{0pt}{2.5ex}                &                    &\lOnecal>20~\GeV\ & \nchrecHLT$\geq$90 &\lOnefcal>50~\GeV&\nchrecHLT$\geq$150 \tabularnewline
\rule{0pt}{2.5ex}                &                    &                &                    &\lOnefcal>50~\GeV&\nchrecHLT$\geq$180 \tabularnewline
\rule{0pt}{2.5ex}                &                    &                &                    &\lOnefcal>65~\GeV&\nchrecHLT$\geq$200 \tabularnewline
\rule[-1.5ex]{0pt}{4.0ex}        &                    &                &                    &\lOnefcal>65~\GeV&\nchrecHLT$\geq$225 \tabularnewline\hline
\end{tabular}
\par\end{centering}
\caption{
The list of L1 and \nchrecHLT requirements for 
  the \pp and \pPb \HMT triggers used in this analysis.
For the \pp HMT triggers, the L1 requirement is on the \et over the entire 
  ATLAS calorimetry (\lOnecal) or hits in the MBTS.
For the \pPb \HMT triggers, the L1 requirement is on the \et restricted 
  to the FCal (\lOnefcal).
\label{tab:pPb_HMT_triggers}
}
\end{table}

\section{Data sets}
\label{sec:data}
The $\sqs = 13$ and 5.02~\TeV\ \pp\ data 
  were collected during Run\,2 of the LHC.
The 13~\TeV\ \pp data were recorded over two periods: a set of low-luminosity
  runs in June 2015 (used in Ref.\,\cite{HION-2015-09}) 
  for which the number of collisions per bunch crossing, $\mu$, varied between
  $0.002$ and $0.04$, and a set of intermediate-luminosity runs in 
  August 2015 where $\mu$ varied between 0.05 and 0.6.
The 5.02~\TeV\ \pp data were recorded during November 2015 in 
  a set of intermediate-luminosity runs with $\mu$ of $\sim$1.5.
The \pPb data were recorded in Run\,1 during \pPb\ operation of the LHC in January 2013.
During that period, the LHC was configured with a 4~\TeV\ proton beam and a 1.57~\TeV\ per-nucleon 
  Pb beam that together produced collisions at \snn=5.02~\TeV.
The higher energy of the proton beam produces a net rapidity shift of the 
  nucleon–nucleon center-of-mass frame by 0.47 units in the proton-going direction,
  relative to the ATLAS reference system. The \pPb\ data were collected in
  two periods between which the directions of the proton and lead beams
  were reversed. 
The integrated luminosities for the three datasets are as follows: \lintaa for the 
   $\sqs{=}13$~\TeV\ \pp data, \lintbb for the $\sqs{=}5.02$~\TeV\ \pp\ data and \lintcc 
   for the $\snn{=}5.02$~\TeV\ \pPb\ data.
However due to the large interaction rates, the full luminosities could not be 
   sampled by the various HMT Triggers listed in Table\,\ref{tab:pPb_HMT_triggers}.
In the $\sqs{=}13$~\TeV\ and $\sqs{=}5.02$~\TeV\ \pp data, the luminosity sampled 
   by the HMT trigger with the highest \lOnecal\ and  \nchrecHLT thresholds 
   were \linta and \lintb, respectively.
In the $\snn{=}5.02$~\TeV\ \pPb\ data, the $\nchrecHLT\geq$225 trigger sampled the 
   entire \lintc luminosity.

\subsection{Event and track selection \label{sec:event_and_track_selection}}
In the offline analysis, additional requirements are imposed on the events 
  selected by the MB and \HMT\ triggers.
The events are required to have a reconstructed vertex with the $z$-position
  of the vertex restricted to $\pm$150~mm.
In the \pPb data, non-collision backgrounds are suppressed by requiring 
  at least one hit in a MBTS counter on each side of the interaction point, 
  and the time-difference measured between the two sides of the MBTS 
  to be less than 10~ns.
In the 2013 \pPb run, the luminosity conditions provided by the LHC resulted 
  in an average probability of 3\% for pileup events.
The pileup events are suppressed by rejecting events containing more than 
  one good reconstructed vertex.
The remaining pileup events are further suppressed using the number of
  detected neutrons, $N_n$, measured in the ZDC on the Pb-fragmentation side.
The distribution of $N_n$ in events with pileup is broader than that for the 
  events without pileup.
Hence, rejecting events at the high tail end of the ZDC signal distribution 
  further suppresses the pileup, while retaining more than 98\% of the 
  events without pileup. 
In the \pp data, pileup is suppressed by only using tracks 
  associated with the vertex having the largest $\sum\pt^2$, where the sum
  is over all tracks associated with the vertex.
Systematic uncertainties in the measured \vn associated with the residual pileup
are estimated in Section\,\ref{sec:systematics}.

In the \pPb analysis, charged-particle tracks are reconstructed in the ID 
  using an algorithm optimized for \pp minimum-bias measurements~\cite{STDM-2010-06}. 
The tracks are required to have \pT>0.4~\GeV\ and $|\eta|$<2.5, at least 
  one pixel hit, with the additional requirement of a hit in the first 
  pixel layer when one is expected,\footnote{A hit is expected if the extrapolated track crosses an active
            region of a pixel module that has not been disabled.}
  and at least six SCT hits.
In addition, the transverse ($d_0$) and longitudinal ($z_0\sin(\theta)$) 
  impact parameters of the track relative to the vertex are required 
  to be less than 1.5~mm. 
They are also required to satisfy $|d_0|/\sigma_{d_0}$<3 and 
  $|z_{0}\sin(\theta)|/\sigma_{z_{0}\sin(\theta)}$<3, where $\sigma_{d_0}$ and $\sigma_{z_{0}\sin(\theta)}$ 
  are uncertainties in $d_0$ and $z_{0}\sin(\theta)$, respectively.

In the \pp analysis, charged-particle tracks and primary vertices are 
  reconstructed in the ID using an algorithm similar to that used in 
  Run\,1, but substantially modified to improve 
  performance~\cite{Salzburger:2018442,ATL-PHYS-PUB-2015-006}.
The reconstructed tracks are required to satisfy the following selection 
  criteria: $\pT$>0.4~\GeV\ and $|\eta|$<2.5;
  at least one pixel hit, with the additional requirement of a hit in 
  the IBL if one is expected (if a hit is not expected in the IBL, a 
  hit in the next pixel layer is required if such a hit is expected); 
  a minimum of six hits in the SCT; 
  $|d_{0}|$<1.5~mm and $|z_0\sin(\theta)|$<1.5~mm.\footnote{In the \pp analysis
  the transverse impact parameter $d_{0}$ is calculated with respect to the 
  average beam position, and not with respect to the vertex.} 
Finally, in order to remove tracks with mismeasured \pT\ due to interactions with the
  material or other effects, the track-fit $\chi^2$ probability is required to be
  larger than 0.01 for tracks having $\pT > 10$~\GeV.

The efficiencies $\epsilon(\pT, \eta)$ of track reconstruction for the
  above track selection cuts are obtained using Monte Carlo (MC) generated
  events that are passed through a GEANT4~\cite{Agostinelli:2002hh} 
  simulation~\cite{SOFT-2010-01} of the ATLAS detector response and 
  reconstructed using the algorithms applied to the data. 
For determining the \pPb efficiencies, the events are generated with 
  version 1.38b of the HIJING event generator~\cite{Wang:1991} with a 
  center-of-mass boost matching the beam conditions.
For determining the \pp efficiencies, non-diffractive 13~\TeV\ \pp events 
  obtained from the \PYEight~\cite{Sjostrand:2007gs} event generator
  (with the A2 set of tuned parameters~\cite{ATL-PHYS-PUB-2011-009} and the MSTW2008LO PDFs~\cite{Sherstnev:2007nd})
  are used.
Both the \pp and \pPb efficiencies increase by $\sim$3\% from 0.4~\GeV\ to 0.6~\GeV\ 
  and vary only weakly with \pt for $\pT$>0.6~\GeV.
In the \pPb case, the efficiency at $\pt\sim$0.6~\GeV\ ranges from 81\% at 
  $\eta$=0 to 73\% at $|\eta|$=1.5 and 65\% at $|\eta|$>2.0. 
The efficiency is also found to vary by less than 2\% over the 
  multiplicity range used in the analysis. 
In the \pp case, the efficiency at $\pt\sim$0.6~\GeV\ ranges from 87\% 
  at $\eta$=0 to 76\% at $|\eta|$=1.5 and 69\% for $|\eta|$>2.0.

\subsection{Event-activity classes}
As in previous ATLAS analyses of long-range correlations in \pPb~\cite{HION-2012-13,HION-2013-04} 
  and \pp~\cite{HION-2015-09} collisions,
  the event activity is quantified by \nchrec: the total number of reconstructed
  charged-particle tracks with $\pt{>}0.4$~\GeV, passing the track selections
  discussed in Section\,\ref{sec:event_and_track_selection}.
From the simulated events (Section\,\ref{sec:event_and_track_selection}), 
  it is determined that the tracking efficiency reduces the measured 
  \Ntrk relative to the event generator multiplicity for \pt>0.4~\GeV\ primary
  charged particles\footnote{For the $p$+Pb simulation, the event generator 
  multiplicity includes charged particles that originate directly from the
  collision or result from decays of particles with $c\tau{<}10$\,mm.   
  The definition for primary charged particles is somewhat tighter in the
  $pp$ simulation~\cite{ATLAS:2016mok}.} by approximately multiplicity-independent factors.
The reduction factors and their uncertainties are
  $1.29\pm0.05$ and $1.18\pm0.05$ for the \pPb and \pp collisions, respectively.

For \pPb\ collisions there is a direct correlation between \nchrec and the 
  number of participating nucleons in the Pb nucleus: events with larger
  \nchrec values have, on average, a larger number of participating nucleons 
  in the Pb nucleus and a smaller impact parameter.
In this case, the concept of centrality used in \NucNuc\ collisions is applicable,
  and in this \papertype\ the terms ``central'' and ``peripheral'' are 
  used to refer to events with large and small values of \nchrec, respectively.
For  \pp\ collisions there may not be a correlation between \nchrec\
  and impact parameter. However, for convenience, the \pp 
  events with large and small \nchrec are also termed as ``central'' 
  and ``peripheral'', respectively. 

\begin{figure}
\includegraphics[width=1.0\linewidth]{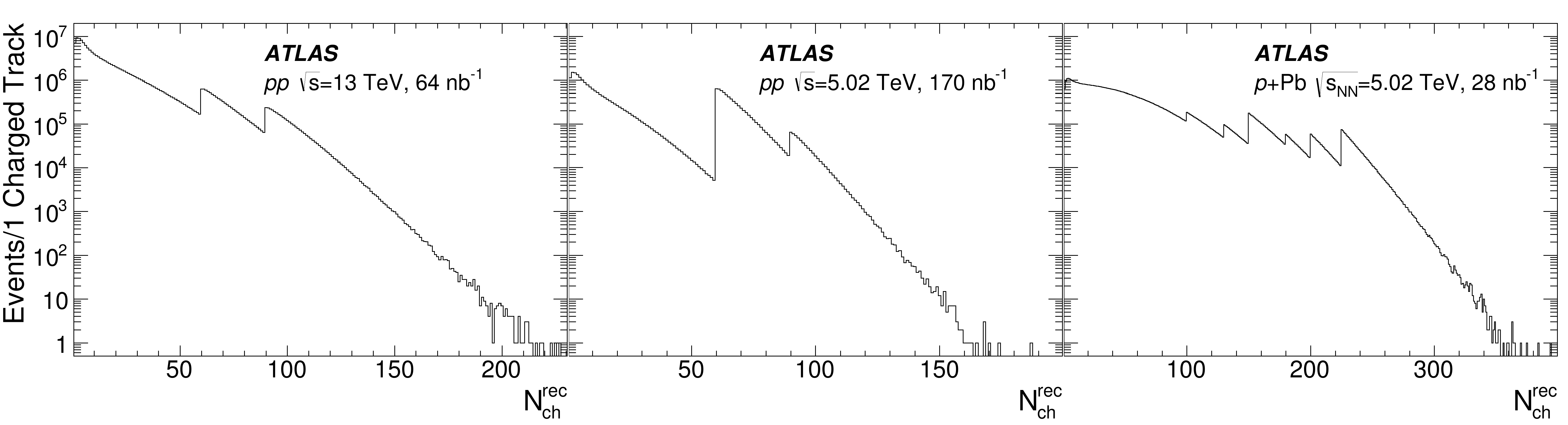}
\caption{Distributions of the multiplicity, $\nchrec$, of reconstructed charged
 particles having $\pt$ >0.4 GeV in the 13~\TeV\ \pp (left), 5.02~\TeV\ \pp (middle), 
 and 5.02~\TeV\ \pPb (right) data used in this analysis.
The discontinuities in the distributions correspond to different high-multiplicity
 trigger thresholds.
\label{fig:nchrec_distributions}}
\end{figure}

Figure\,\ref{fig:nchrec_distributions} shows the $\nchrec$ distributions
  for the three data sets used in this \papertype.
The discontinuities in the distributions result from the different \HMT
  triggers, for which an offline requirement of \nchrec>\nchrecHLT 
  is applied. 
This requirement ensures that the \HMT triggered events are used only where the HLT 
  trigger is almost fully efficient.

The \pp\ event activity can also be quantified using the
  total transverse energy deposited in the FCal (\fcalet).  
This quantity has been used to determine the centrality in all ATLAS
  heavy-ion analyses.
Using the \fcalet to characterize the event activity has the advantage 
  that independent sets of particles are used to determine the 
  event activity and to measure the long-range correlations.  
Similarly in the \pPb case, the event activity can be characterized by 
  the sum of transverse energy measured on the Pb-fragmentation side of
  the FCal (\fcaletPb)~\cite{HION-2012-13,HION-2013-04}.
Results presented in this \papertype\ use both \nchrec and the \fcalet
  (or \fcaletPb) to quantify the event activity.

\section{Two-particle correlation analysis}
\label{sec:corr}
The study of two-particle correlations in this \papertype\ follows
   previous ATLAS measurements in \PbPb~\cite{HION-2011-01,HION-2012-10,HION-2014-03}, 
   \pPb~\cite{HION-2012-13,HION-2013-04} and \pp~\cite{HION-2015-09} collisions.
For a given event class, the two-particle correlations are measured 
  as a function of the relative azimuthal angle $\dphi\equiv\phi^a-\phi^b$ 
  and pseudorapidity $\deta \equiv\eta^a-\eta^b$ separation. 
The labels  $a$ and $b$ denote 
  the two particles in the pair, which may be selected from different 
  \pT intervals. 
The particles $a$ and $b$ are conventionally referred to as the ``trigger'' 
  and ``associated'' particles, respectively. 
The correlation function is defined as:
\begin{eqnarray} 
\label{eq:ana0}
\ctwo =\frac{\stwo}{B(\deta,\dphi)}\;,
\end{eqnarray}
where $S$ and $B$ represent pair distributions constructed from the
  same event and from ``mixed events''~\cite{Adare:2008ae}, 
  respectively. 
The same-event distribution $S$ is constructed using all particle pairs that 
  can be formed in each event from tracks that have passed the
  selections described in Section\,\ref{sec:event_and_track_selection}. 
The $S$ distribution contains both the physical correlations between particle pairs and
  correlations arising from detector acceptance effects.
The mixed-event distribution $B(\deta,\dphi)$ is similarly constructed by choosing 
  the two particles in the pair from different events. 
The $B$ distribution does not contain physical correlations, but has detector acceptance effects 
  similar to those in $S$. 
In taking the ratio, $S/B$ in Eq.\,\eqref{eq:ana0}, the detector acceptance effects
  largely cancel, and the resulting \ctwo contains physical correlations only.
The pair of events used in the mixing are required to have similar 
  $\Ntrk$ ($|\Delta\Ntrk|{<}10$) and similar $\zvtx$ ($|\Delta\zvtx|{<}10$~mm), 
  so that acceptance effects in $\stwo$ are properly reflected in, and 
  compensated by, corresponding variations in $B(\deta,\dphi)$. 
To correct $\stwo$ and $B(\deta,\dphi)$ for the individual
  $\phi$-averaged inefficiencies of particles $a$ and $b$, the pairs 
  are weighted by the inverse product of their tracking efficiencies
  $1/(\epsilon_a\epsilon_b)$. 
Statistical uncertainties are calculated for \ctwo\ using standard
  error-propagation procedures assuming no correlation between $S$ and
  $B$, and with the statistical variance of $S$ and
  $B$ in each \deta\ and \dphi\ bin taken to be $\sum
  1/(\epsilon_a\epsilon_b)^2$ 
  where the sum runs over all of the pairs included in the bin.
Typically, the two-particle correlations are used only to study the shape
  of the correlations in \dphi, and are conveniently normalized.
In this \papertype, the normalization of \ctwo\ is chosen such that the 
  \dphi-averaged value of \ctwo\ is unity for $|\deta|>2$.

\begin{figure}
\includegraphics[width=0.9\linewidth]{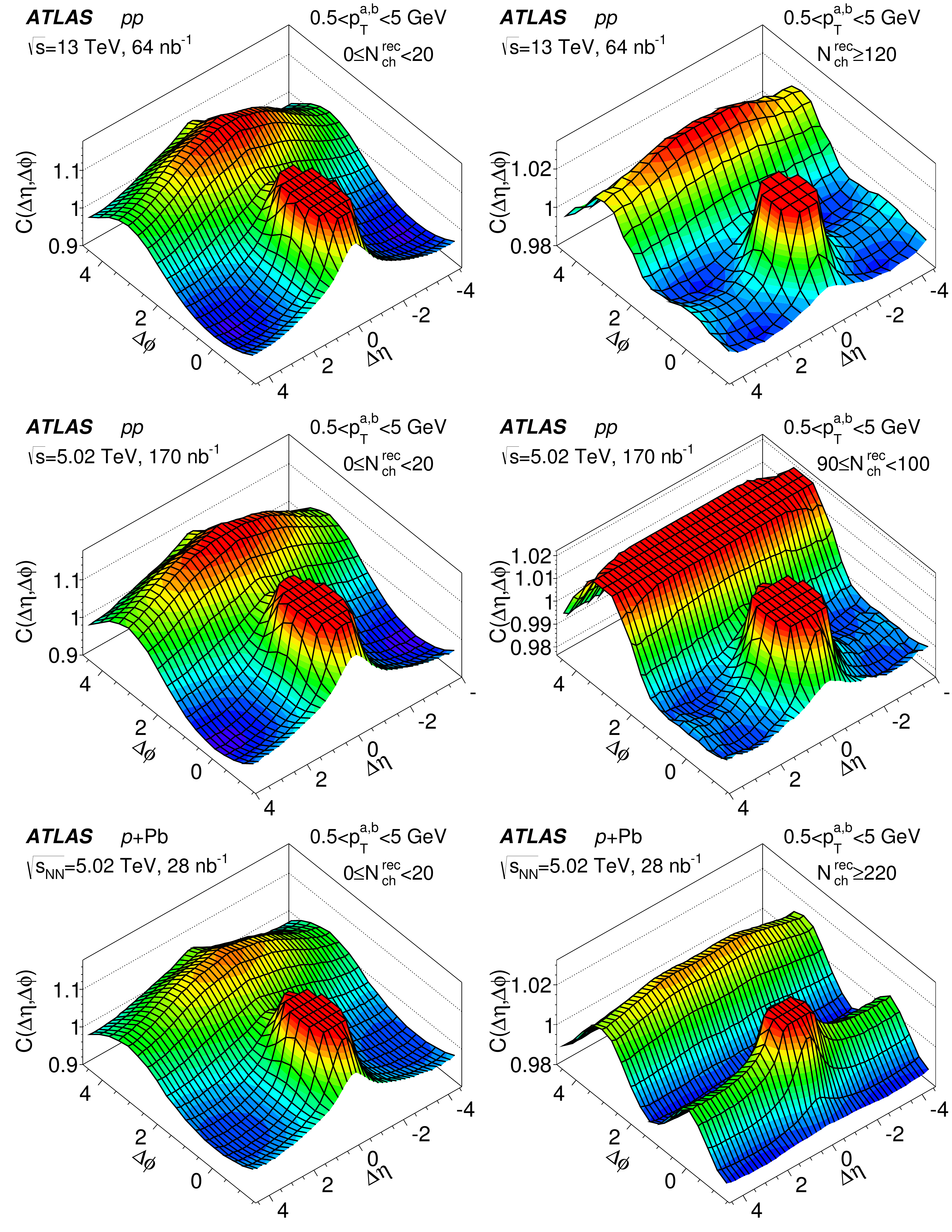}
\caption{
Two-particle correlation functions \ctwo in 13~\TeV\ \pp collisions
  (top panels), 5.02~\TeV\ \pp collisions (middle panels) and in 5.02~\TeV\
  \pPb\ collisions (bottom panels). 
The left panels correspond to a lower-multiplicity range of 
  $0{\leq}\nchrec{<}20$.
The right panels correspond to higher multiplicity ranges of 
  $\nchrec{\geq}120$ for 13~\TeV\ \pp, $90{\leq}\nchrec{<}100$ for the 5.02~\TeV\ \pp\
  and $\nchrec{\geq}220$ for the  5.02~\TeV\ \pPb. 
The plots are for charged particles having 0.5<\ptab<5~\GeV. 
The distributions have been truncated to suppress the peak at 
  $\deta{=}\dphi{=}0$ and are plotted over $|\deta|{<}4.6$ 
  ($|\deta|{<}4.0$ for middle row) to avoid 
  statistical fluctuations at larger $|\deta|$. 
For the middle-right panel, the peak at $\dphi{=}\pi$ has also been truncated.
\label{fig:2d_corrs}}
\end{figure}

Examples of correlation functions are shown in Figure\,\ref{fig:2d_corrs} for
  $0.5{<}\ptab{<}5$~\GeV\ and for two different 
  \Ntrk\ ranges for each of the three data sets: 13~\TeV\ \pp\
  (top), 5.02~\TeV\ \pp (middle), and 5.02~\TeV\ \pPb\ (bottom).
The left panels show results for $0{\leq}\nchrec{<}20$ while the right panels show 
  representative high-multiplicity ranges
  of $\nchrec{\geq}120$ for the 13~\TeV\ \pp data, $90{\leq}\nchrec{<}100$ for the 5.02~\TeV\ \pp data
  and $\nchrec{\geq}220$ for the 5.02~\TeV\ $\pPb$ data.
The correlation functions are plotted over the range $-\pi/2{<}\dphi{<}3\pi/2$; 
  the periodicity of the measurement requires that
  $C(\deta, 3\pi/2) {=} C(\deta, -\pi/2)$. 
The low-multiplicity correlation functions exhibit features that are 
  understood to result primarily from hard-scattering processes: 
  a peak centered at $\deta{=}\dphi{=}0$ that arises primarily from jets 
  and an enhancement centered at $\dphi{=}\pi$ and extending over the 
  full \deta\ range which results from dijets. 
These features also dominate the high-multiplicity correlation
  functions. 
Additionally, in the high-multiplicity correlation functions, 
  each of the three systems exhibit a ridge -- an enhancement 
  centered at $\dphi {=} 0$ that extends over the entire measured \deta\ range.

One-dimensional correlation functions \ctwophi are obtained by
  integrating the numerator and denominator of Eq.\,\eqref{eq:ana0} over
  2<|{\deta}|<5 prior to taking the ratio:
\begin{eqnarray} 
\ctwophi = \frac{\int_{2}^{5} d|\deta| \; \stwoabs}{\int_{2}^{5}
          d|\deta| \; B(|\deta|,\dphi)}\equiv\frac{\stwophi}{B(\dphi)}.
\end{eqnarray}
This |{\deta}|\ range is chosen to focus on the long-range features 
  of the correlation functions. 
From the one-dimensional correlation functions, ``per-trigger-particle yields,''
 \Yphi\ are calculated~\cite{Adare:2008ae,HION-2012-13,HION-2013-04}:
\begin{eqnarray}
\label{eq:pty}
\Yphi  =\left( \frac{\int_{-\pi/2}^{3\pi/2} B(\dphi) d\dphi}{ N^{a} \int_{-\pi/2}^{3\pi/2} d\dphi}  \right) \ctwophi,
\end{eqnarray}
where $N^a$ denotes the total number of trigger particles, corrected to account 
  for the tracking efficiency.
The \Yphi distribution is identical in shape to \ctwophi, but has a physically relevant 
  normalization: it represents the average number of associated particles 
  per trigger particle in a given \dphi interval. 
This allows operations, such as subtraction of the \Yphi distribution in one event-activity 
  class from the \Yphi distribution in another, which have been used in studying the
   \pPb ridge~\cite{HION-2012-13,HION-2013-04}.

\section{Template fitting}
\label{sec:fits}

In order to separate the ridge from other sources of angular correlation, 
  such as dijets, the ATLAS Collaboration  developed a template fitting procedure 
  described in Ref.\,\cite{HION-2015-09}.
In this procedure, the measured \Yphi\ distributions are 
  assumed to result from a superposition of a ``peripheral'' \Yphi distribution, \YphiPer,
  scaled up by a multiplicative factor and a constant modulated by $\cos(n\dphi)$
  for $n\geq$2. 
The resulting template fit function,
\begin{equation}
\YphiTempl  = \YphiRidge + F \, \YphiPer   \, ,
\label{eq:template}
\end{equation}
where
\begin{equation}
\YphiRidge = G \left(1 + \sum_{n=2}^{\infty}2\vnn \cos{(n\Delta \phi)}\right)\, ,
\label{eq:template_ridge}
\end{equation}
has free parameters $F$ and \vnn. 
A $v_{1,1}$ term is not included in \YphiRidge (Eq.\,\eqref{eq:template_ridge})
  as the presence of a $v_{1,1}$ component in the measured $Y(\Delta\phi)$
  is accounted for by the $F\YphiPer$ term.
The parameter $F$ is the multiplicative factor by which the \YphiPer is scaled.
The coefficient $G$, which represents the magnitude of the combinatoric 
  component of \YphiRidge, is fixed by requiring that 
  the integral of \YphiTempl be equal to the integral of the measured \Yphi:
  $\int_0^{\pi}{d\dphi}\; \YphiTempl = \int_0^{\pi}{d\dphi} \;\Yphi$. 
In this \papertype, when studying the \Ntrk dependence of the long-range correlation, 
  the $0{\leq}\Ntrk{<}20$ multiplicity interval is used to produce \YphiPer.
When studying the \fcalet (\fcaletPb) dependence, the \fcalet<10~\GeV\ 
  (\fcaletPb<10~\GeV) interval is used to produce \YphiPer.

The template fitting procedure is similar to the peripheral subtraction 
  procedure used in previous ATLAS \pPb ridge analyses~\cite{HION-2013-04}.
In those analyses, the scale factor for the peripheral reference, 
  analogous to $F$ in Eq.\,\eqref{eq:template}, was determined by matching the 
  near-side jet peaks between the peripheral and central samples.
A more important difference, however, lies in the treatment of the peripheral bin.
In the earlier analyses, a ZYAM procedure was performed on the peripheral reference,
  and only the modulated part of \YphiPer, $\YphiPer-\YphiPerZ$, was used in the 
  peripheral subtraction.\footnote{The minimum of \YphiPer is at
   $\dphi{=}0$ and is thus equal to \YphiPerZ, which the ZYAM procedure subtracts out.}
The ZYAM procedure makes several assumptions, the most relevant of which for 
  the present analysis is that there is no long-range correlation in the peripheral bin.
As pointed out in Ref.\,\cite{HION-2015-09}, neglecting the non-zero
  modulation present in \YphiPer\ significantly biases the measured \vnn values.
Results from an alternative version of the template fitting, where a 
  ZYAM procedure is performed on the peripheral reference, by using $\YphiPer-\YphiPerZ$
  in place of \YphiPer in Eq.\,\eqref{eq:template}, are also presented 
  in this \papertype.
This ZYAM-based template fit is similar to the \pPb peripheral subtraction procedure.
These results are included mainly to compare with previous measurements 
  and to demonstrate the improvements obtained using the present method.

\begin{figure}
\begin{centering}
\includegraphics[width=1.0\linewidth]{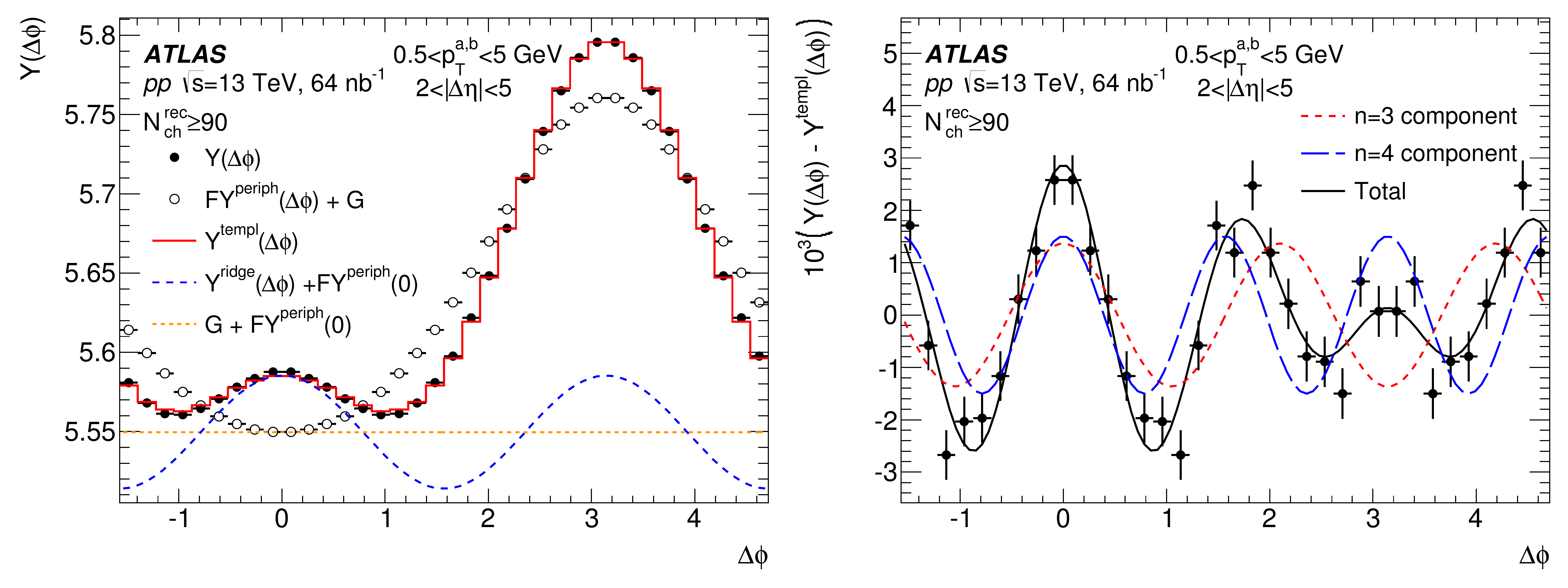}
\end{centering}
\vspace{-0.5cm}
\caption{
Left Panel: template fit to the per-trigger particle yields \Yphi
  in 13~\TeV\ \pp collisions for charged-particle pairs with $0.5{<}\ptab{<}5$~\GeV\
  and $2{<}|\deta|{<}5$. 
This plot corresponds to the $\nchrec{\geq}90$ multiplicity range. 
The template fitting includes only the second-order harmonic, \vtt. 
The solid points indicate the measured \Yphi, the open points and
  curves show different components of the template (see legend)
  that are shifted along the $y$-axis by $G$ or by $F\YphiPerZ$, where necessary, for presentation. 
Right Panel: The difference between the $\Yphi$ and the template
  fit, showing the presence of \vthth and \vff components. 
The vertical error bars indicate statistical uncertainties.
\label{fig:fits_v2_only}}
\end{figure}

In Ref.\,\cite{HION-2015-09} the template fitting procedure only 
  included the second-order harmonic \vtt, but was able to 
  reproduce the \nchrec-dependent evolution of \Yphi on both the 
  near and away sides.
The left panel of Figure\,\ref{fig:fits_v2_only} shows such a template fit,
  in the 13~\TeV\ \pp data, that only includes \vtt.
The right panel shows the difference between the \Yphi and the \YphiTempl distributions
  demonstrating the presence of small (compared to \vtt), but significant 
  residual \vthth and \vff components. 
While it is possible that $\cos{3\dphi}$ and $\cos{4\dphi}$ contributions 
  could arise in the template fitting method due to small multiplicity-dependent 
  changes in the shape of the dijet component of the correlation function, such
  effects would not produce the excess at $\dphi {\sim} 0$ observed in
  the right-hand panel in Figure\,\ref{fig:fits_v2_only}. 
That excess and the fact that its magnitude is compatible with the remainder of the
  distribution indicates that there is real $\cos{3\dphi}$ and $\cos{4\dphi}$ 
  modulation in the two-particle correlation functions. 
Thus this \papertype\ extends the \vtt results in Ref.\,\cite{HION-2015-09} 
  by including \vthth and \vff as well.
A study of these higher-order harmonics, including their \nchrec and \pt dependence 
  and factorization (Eq.\,\eqref{eq:factortwo}), can help in better understanding
  the origin of the long-range correlations.

\begin{figure}
\begin{centering}
\includegraphics[width=0.98\linewidth]{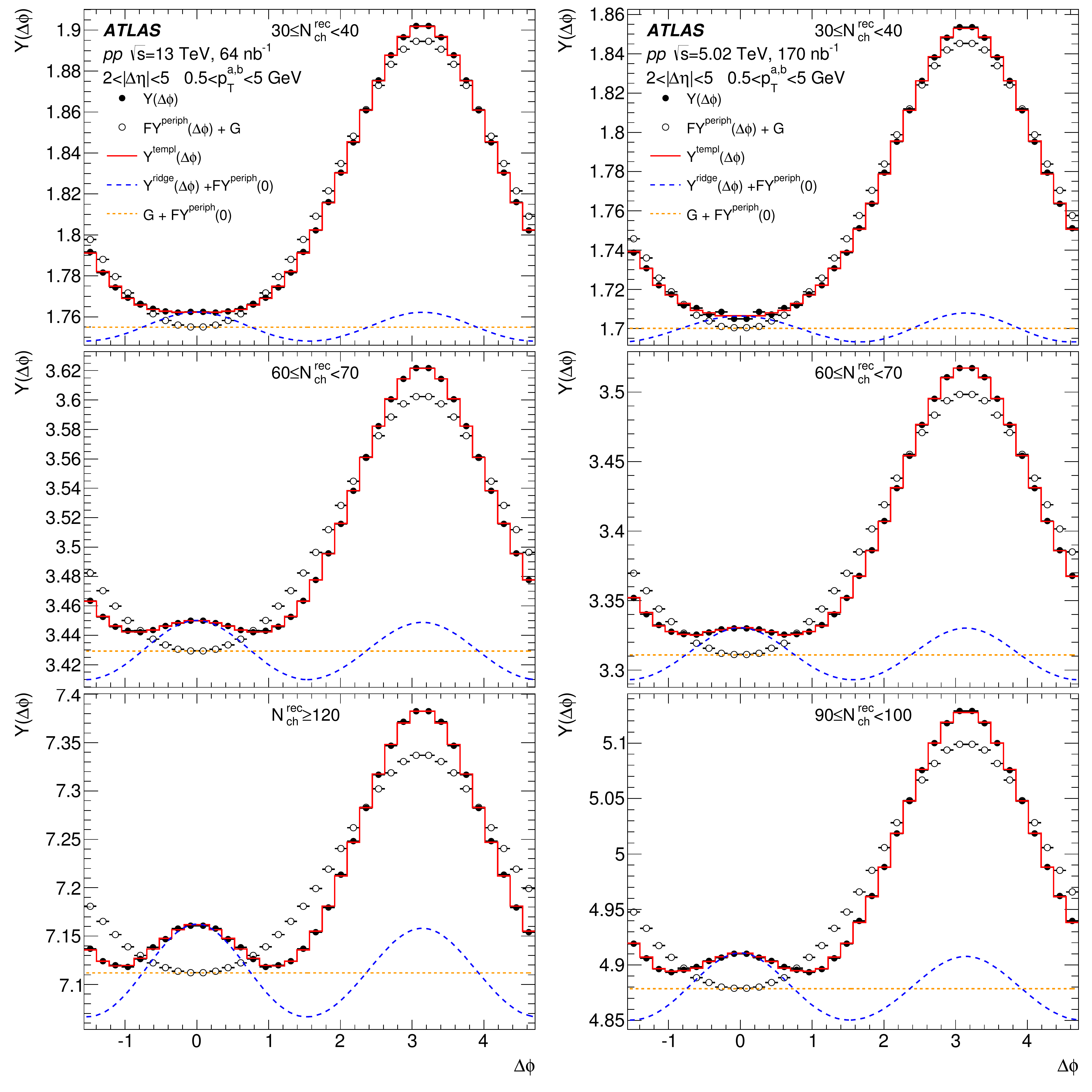}
\end{centering}
\vspace{-0.3cm}
\caption{
Template fits to the per-trigger particle yields \Yphi, in 13~\TeV\
  (left panels) and in 5.02~\TeV\ (right panels) \pp collisions for
  charged-particle pairs with $0.5{<}\ptab{<}5$~\GeV\ and $2{<}|\deta|{<} 5$.
The template fitting includes second-order, third-order and fourth-order harmonics. 
From top to bottom, each panel represents a different \nchrec range. 
The solid points indicate the measured \Yphi, the open points and
  curves show different components of the template (see legend)
  that are shifted along the $y$-axis by $G$ or by $F\YphiPerZ$, where necessary, for presentation. 
\label{fig:fits_pp}}
\end{figure}

Figure~\ref{fig:fits_pp} shows template fits to the 13~\TeV\ (left panels)
  and 5.02~\TeV\ \pp data (right panels), for $0.5<\ptab<5$~\GeV. 
From top to bottom, each panel represents a different \nchrec range. 
The template fits (Eq.\,\eqref{eq:template_ridge}) include harmonics 2--4.
Visually, a ridge, i.e. a peak on the near-side,
  cannot be seen in the top two rows, which correspond to low and 
  intermediate \nchrec intervals, respectively.
However, the template fits indicate the presence of a large modulated
  component of \YphiRidge even in these \nchrec intervals. 
Several prior \pp ridge measurements rely on the ZYAM 
  method~\cite{Ajitanand:2005jj,Adare:2008ae} to extract  
  yields on the near-side~\cite{Khachatryan:2010gv,CMS:2012qk}.
In these analyses, the yield of excess pairs in the ridge above the
  minimum of the \Yphi distribution is considered to be the strength of the ridge.
Figure\,\ref{fig:fits_pp} shows that such a procedure would
  give zero yields in low- and intermediate-multiplicity collisions
  where the minimum of \Yphi occurs at $\dphi{\sim}0$. 
In high-multiplicity events the ZYAM-based yields, while non-zero, 
  are still underestimated.

\begin{figure}
\begin{centering}
\includegraphics[width=0.98\linewidth]{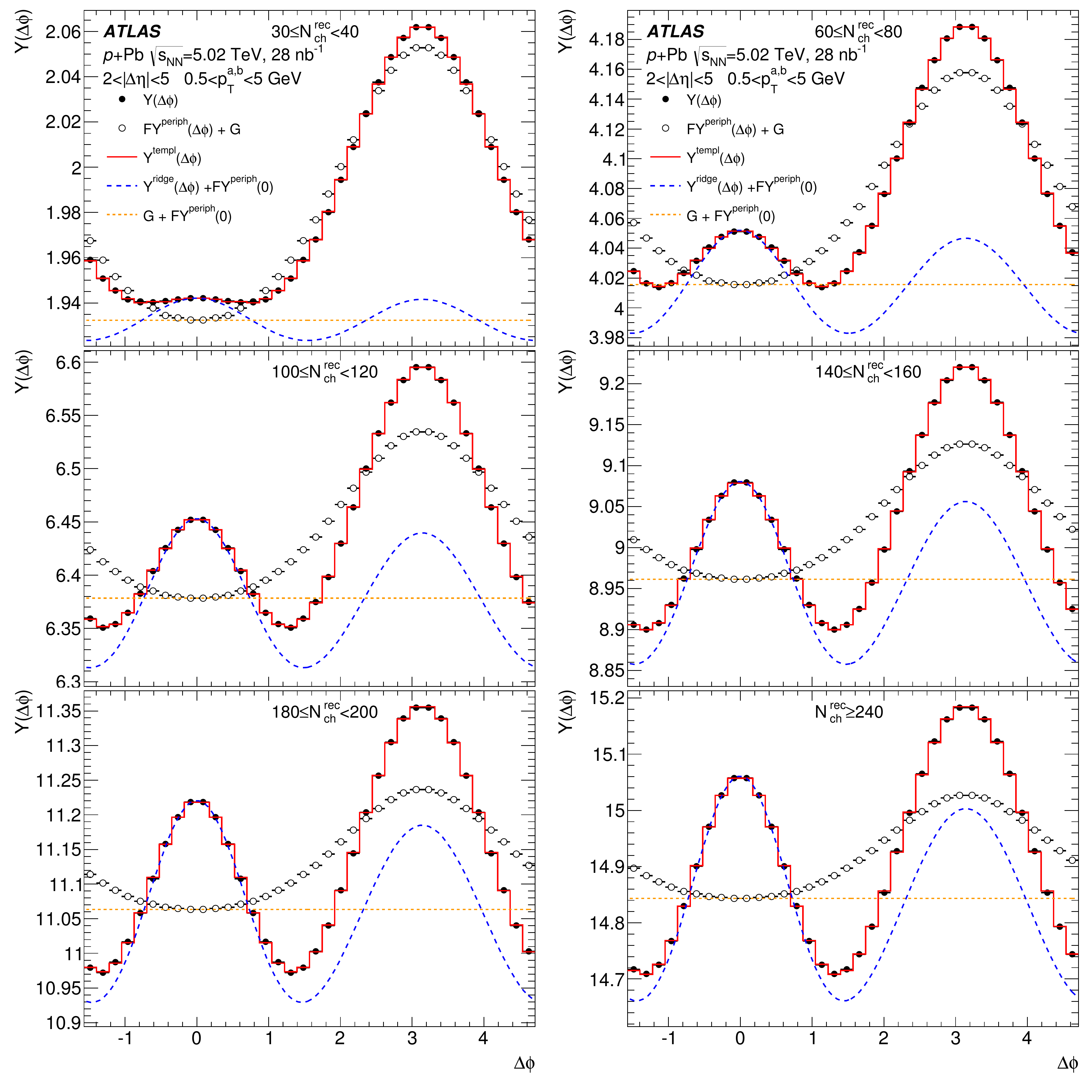}
\end{centering}
\vspace{-0.3cm}
\caption{
Template fits to the per-trigger particle yields \Yphi in 5.02~\TeV\ 
  \pPb\ collisions for charged-particle pairs with $0.5{<}\ptab{<}5$~\GeV\
  and $2{<}|\deta|{<}5$. 
The template fitting includes second-order, third-order and fourth-order harmonics. 
Each panel represents a different \nchrec range. 
The solid points indicate the measured \Yphi, the open points and
  curves show different components of the template (see legend)
  that are shifted along the $y$-axis by $G$ or by $F\YphiPerZ$, where necessary, for presentation. 
\label{fig:fits_pPb}}
\end{figure}

Figure~\ref{fig:fits_pPb} shows the template fits to the \pPb data 
  in a format similar to Figure\,\ref{fig:fits_pp}.
The template fits describe the data well across the entire
  \nchrec range used in this \papertype.
Previous \pPb ridge analyses used a peripheral subtraction procedure
  to remove the jet component from \Yphi~\cite{HION-2012-13,HION-2013-04,
  Abelev:2012ola,Chatrchyan:2013nka,HION-2013-04,Khachatryan:2015waa}.
That procedure is similar to the ZYAM-based template fitting 
  procedure, in that it assumes absence of any long-range correlations 
  in the peripheral events.
In the following sections, comparisons between the \vnn obtained from these 
  two methods are shown.

\clearpage
\subsection{Fourier coefficients}
\begin{figure}
\begin{centering}
\includegraphics[width=0.98\linewidth]{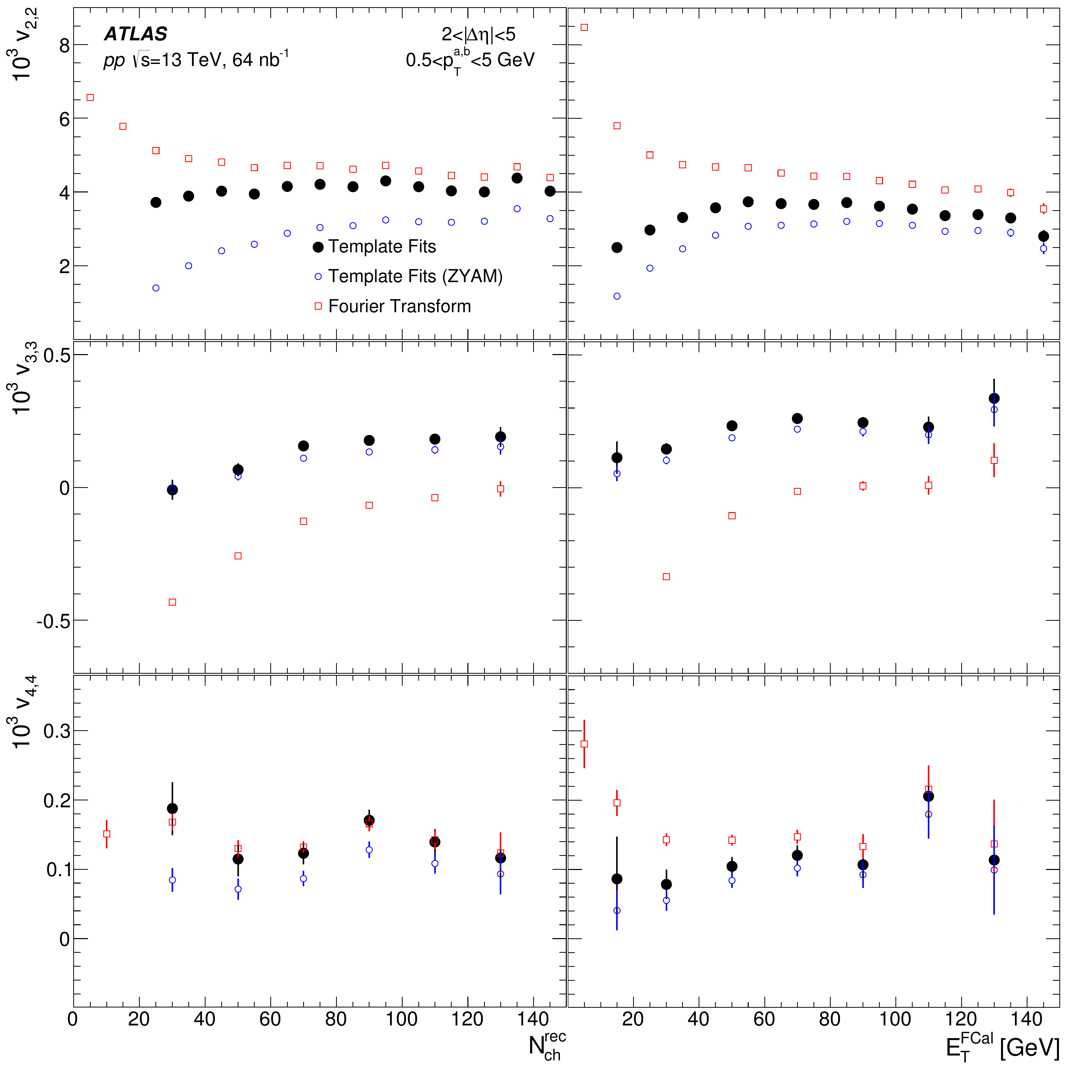}
\vspace{-0.5cm}
\end{centering}
\caption{
The \vnn obtained from the template fitting procedure
  in the 13~\TeV\ \pp data, as a function of \nchrec (left panels),
  and as a function of \fcalet (right panels). 
The top, middle and bottom panels correspond to \vtt, \vthth, and 
  \vff, respectively. 
The results are for $0.5{<}\ptab{<}5$~\GeV. 
The error bars indicate statistical uncertainties.
The \vnn obtained from a direct Fourier transform of \Yphi and
  from the ZYAM-based template fits are also shown for comparison.
\label{fig:fit_parms_13Tev}}
\end{figure}

Figure~\ref{fig:fit_parms_13Tev} shows the \vnn obtained from the template
  fits in the 13~\TeV\ \pp data, as a function of \nchrec and \fcalet.  
The \vnn from the ZYAM-based template fits as well as the coefficients 
  obtained from a direct Fourier transform of \Yphi:
\begin{eqnarray} 
\label{eq:vn_fourier}
\text{Fourier-}\vnn\equiv\frac{\int\Yphi\cos(n\dphi)d\dphi}{\int\Yphi d\dphi}
\end{eqnarray}
 are also shown 
  for comparison.
While the template-\vnn are the most physically meaningful quantities,
  the Fourier-\vnn are also included to demonstrate how the template fitting
  removes the hard contribution.
Similarly, the ZYAM-based template-\vnn are also included, as the 
  ZYAM-based fitting is similar to the peripheral subtraction procedure 
  used in prior \pPb analyses~\cite{HION-2012-13,HION-2013-04}, 
  and comparing with the ZYAM-based results illustrates the improvement
  brought about in the template fitting procedure.

The \vtt values are nearly independent of \nchrec throughout the
  measured range.
As concluded in Ref.\,\cite{HION-2015-09}, this implies that the 
  long-range correlation is not unique to high-multiplicity
  events, but is in fact present even at very low multiplicities.
In the \fcalet dependence, however, \vtt shows a systematic 
  decrease at low \fcalet.
Further, the asymptotic value of the template-\vtt at large \nchrec is also
  observed to be $\sim$10\% larger than the asymptotic value at large \fcalet.
This might indicate that the \vtt at a given rapidity is more
  correlated with the local multiplicity than the global multiplicity.

The removal of the hard-process contribution to \vtt in the template fitting
  can be seen by comparing to the Fourier-\vtt values.
The Fourier-\vtt values are always larger than the template-\vtt and
  show a systematic increase at small \nchrec (\fcalet).
This indicates the presence of a relatively large contribution 
  from back-to-back dijets over this range.
Asymptotically, at large \nchrec the Fourier-\vtt  values become stable,
  but show a small decreasing trend in the \fcalet dependence.
The ZYAM-based \vtt values are smaller than the template-\vtt values for all
  \nchrec (\fcalet), and by construction systematically decrease 
  to zero for the lower \nchrec (\fcalet) intervals.
However, at larger \nchrec (\fcalet) they also show only a weak dependence on 
  \nchrec (\fcalet).
Asymptotically, at large \nchrec the \vtt values from the Fourier transform and 
  the default template fits match to within $\sim$10\% (relative).
In general, the \vtt values from all three methods agree within $\pm$15\% at large
  \nchrec or \fcalet.
This implies that at very high multiplicities, $\nchrec\sim120$, the ridge
  signal is sufficiently strong that the assumptions made in removing
  the hard contributions to \Yphi do not make a large
  difference. 
However, for the highest \pT\ values used in this analysis, 
  $\pta{>}7$~\GeV, it is observed that the width of the dijet peak 
  in the \pp\ correlation functions broadens with increasing multiplicity. 
This change is opposite to that seen at lower \pT\ where \vtt\ causes the dijet
  peak to become narrower. 
As a result, the measured \vtt\ values become negative. 
This bias from the multiplicity dependence of the hard-scattering 
  contribution likely affects the correlation functions at lower 
  \pTab\ values and its potential impact is discussed below.

The \vtt component is dominant, with a magnitude approximately 30 times 
  larger than \vthth and \vff, which are comparable to each other.
This is in stark contrast to \PbPb collisions where in the most central events,
  where the average geometry has less influence,
  the \vnn have comparable magnitudes~\cite{HION-2011-01}.
The Fourier-\vthth shows considerable \nchrec (\fcalet) dependence and
  is negative almost everywhere.
However, the \vthth values from the template fits are mostly positive.
As the factorization of the \vnn requires that
  the \vnn be positive (Eq.\,\eqref{eq:factorone}), the negative 
  Fourier-\vthth clearly does not arise from single-particle 
  modulation.
However, because the template-\vthth is positive, its origin
  from single-particle modulation cannot be ruled out.
Within statistical uncertainties, the \vff values from all three methods are positive throughout the measured \Ntrk range.
Within statistical uncertainties, the \vff values are consistent with no \nchrec 
 or \fcalet dependence.

\begin{figure}
\begin{centering}
\includegraphics[width=0.99\linewidth]{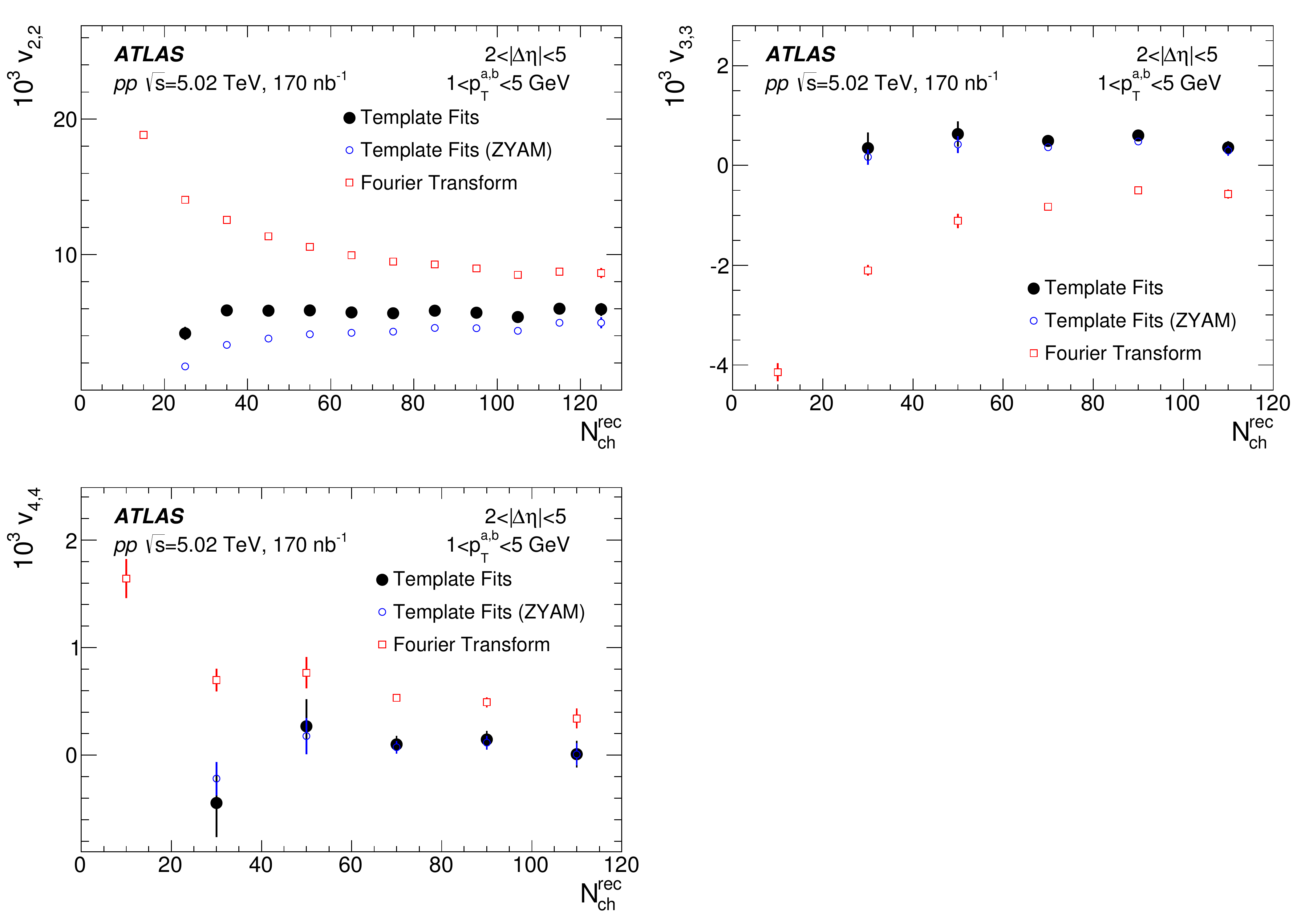}
\end{centering}
\vspace{-0.3cm}
\caption{
The \vnn obtained from the template fitting procedure
  in the 5.02~\TeV\ \pp\ data, as a function of \nchrec.
The three panels correspond to $n=$2, 3, and 4, respectively.
The results are for $1{<}\ptab{<}5$~\GeV. 
The error bars indicate statistical uncertainties.
The \vnn obtained from a direct Fourier transform of \Yphi and
  from the ZYAM-based template fits are also shown for comparison.
\label{fig:fit_parms_5Tev}}
\end{figure}

Figure~\ref{fig:fit_parms_5Tev} shows the \vnn values from the 
  5.02~\TeV\ \pp data as a function of \nchrec for a higher \ptab bin of 
  1--5~\GeV.
The same trends seen in the 13~\TeV\ data
  (Figure\,\ref{fig:fit_parms_13Tev}) are observed here, and the conclusions are 
  identical to those made in the 13~\TeV\ case.

\begin{figure}
\begin{centering}
\includegraphics[width=0.98\linewidth]{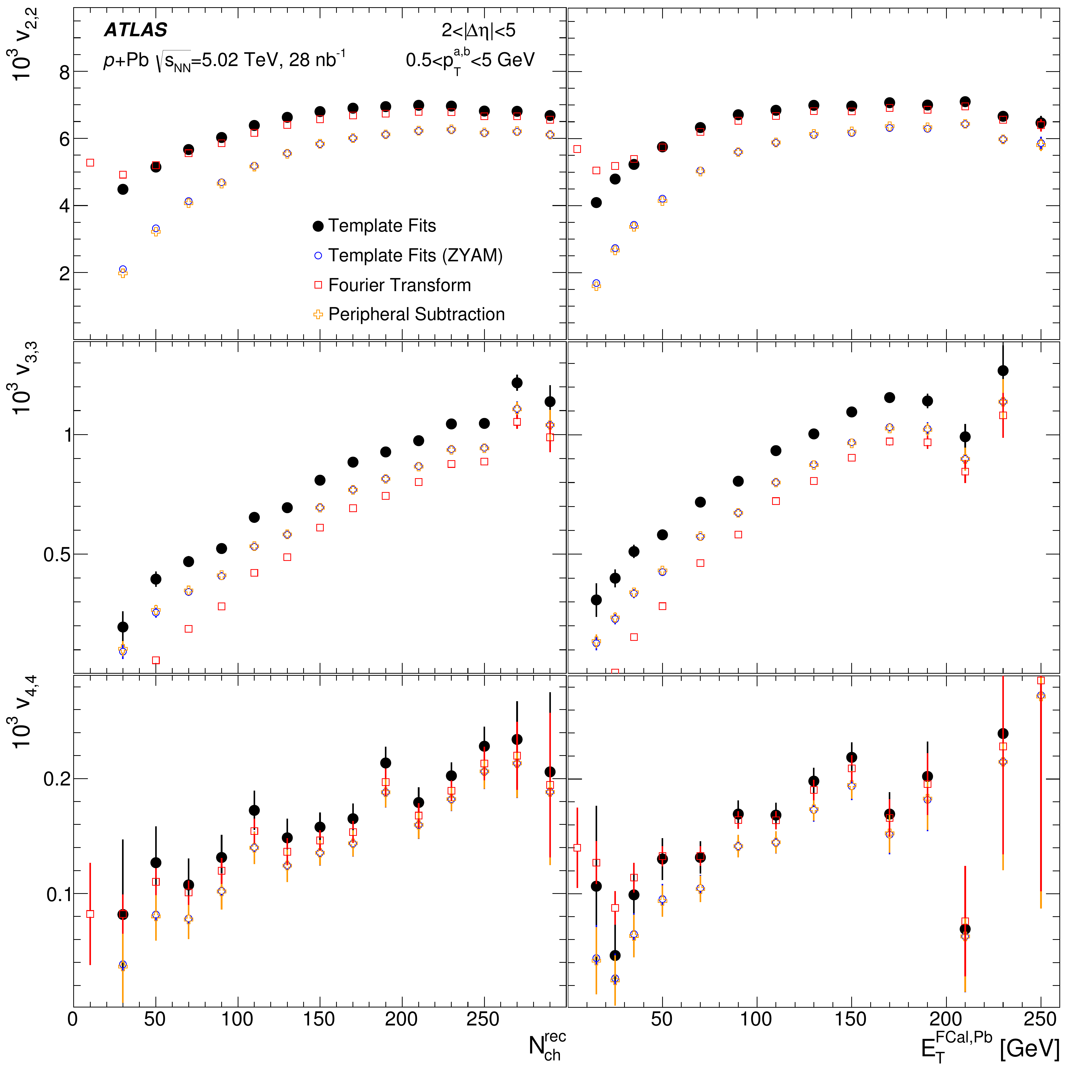}
\end{centering}
\vspace{-0.3cm}
\caption{
The \vnn obtained from the template fitting procedure
  in the 5.02~\TeV\ \pPb data, as a function of \nchrec (left panels),
  and as a function of the Pb-fragmentation side FCal-\et (right panels).
The top, middle and bottom panels correspond to \vtt, \vthth,
  and \vff, respectively. 
The results are for $0.5{<}\ptab<{5}$~\GeV. 
The error bars indicate statistical uncertainties.
The \vnn obtained from a direct Fourier transform of \Yphi,
  the peripheral subtraction procedure, and
  from the ZYAM-based template fits are also shown for comparison.
\label{fig:fit_parms_pPb}}
\end{figure}

Figure~\ref{fig:fit_parms_pPb} shows the \vnn for the \pPb data.
The results are plotted both as a function of \nchrec (left panels)
  and \fcaletPb (right panels).
The \vtt values obtained from the template fits show a systematic increase with
  \nchrec over $\nchrec\lesssim$150, unlike the \pp case where \vtt 
  is nearly independent of \nchrec.
This increase is much larger compared to the systematic uncertainties
  in the \vtt values (discussed later in Section\,\ref{sec:systematics}).
This is possibly indicative of a systematic change in the average collision 
  geometry which is present in \pPb but not in \pp collisions.
A similar increase of the \vtt values is also observed in the \fcaletPb dependence.
The higher-order harmonics \vthth and \vff show a stronger relative increase
  with increasing \nchrec and \fcaletPb.
This also implies that the assumption made in the template-fitting, regarding
  the independence or weak dependence of the \vnn on \nchrec, is not 
  strictly correct for \vthth and \vff.

Figure~\ref{fig:fit_parms_pPb} also compares the Fourier and ZYAM-based 
  template-\vnn values.
The \vnn from the peripheral subtraction procedure used in 
  a previous ATLAS \pPb long-range correlation analysis \cite{HION-2013-04}
  are also shown.
The peripheral-subtracted \vnn values are nearly identical to the
  values obtained from the ZYAM-based template fits.
This is expected, as the treatment of the peripheral bin
  is identical in both cases: both use the ZYAM-subtracted \YphiPer
  as the peripheral reference.
What differs procedurally between the two methods is determination of the 
  scale factor by which \YphiPer is scaled up when subtracting it from \Yphi.
In the peripheral subtraction case, this scale factor, analogous to the 
  parameter $F$ in Eq.\,\eqref{eq:template}, is determined by matching the 
  near-side jet peaks over the region $|\deta|{<}1$ and $|\dphi|{<}1$.
In the template-fitting case, the parameter $F$ is determined by 
  the jet contribution to the away-side peak.
The similarity of the \vtt values from the two procedures implies that whether 
  the matching is done in the near-side jet peak, or over the away-side 
  peak, identical values of the scale factor are obtained.
The Fourier-\vtt and template-\vtt values are surprisingly similar
  except at very low \nchrec or \fcaletPb.
This is unlike the \pp case (Figures~\ref{fig:fit_parms_13Tev} and 
  \ref{fig:fit_parms_5Tev}), where the values differed by 
  $\sim$15\% (relative) at large \nchrec.
This similarity does not hold for \vthth where the values from the 
  template fit are systematically larger than the values obtained 
  from Fourier decomposition.
For all harmonics, the relative difference in the \vnn decreases with
  increasing event activity.
Like in the \pp case (Figure\,\ref{fig:fit_parms_13Tev}), this implies
  that at large enough event activity, the \vnn are less sensitive to
  the assumptions made in removing the hard contributions.

\subsection{Test of factorization in template fits}

\begin{figure}
\begin{centering}
\includegraphics[width=1.0\linewidth]{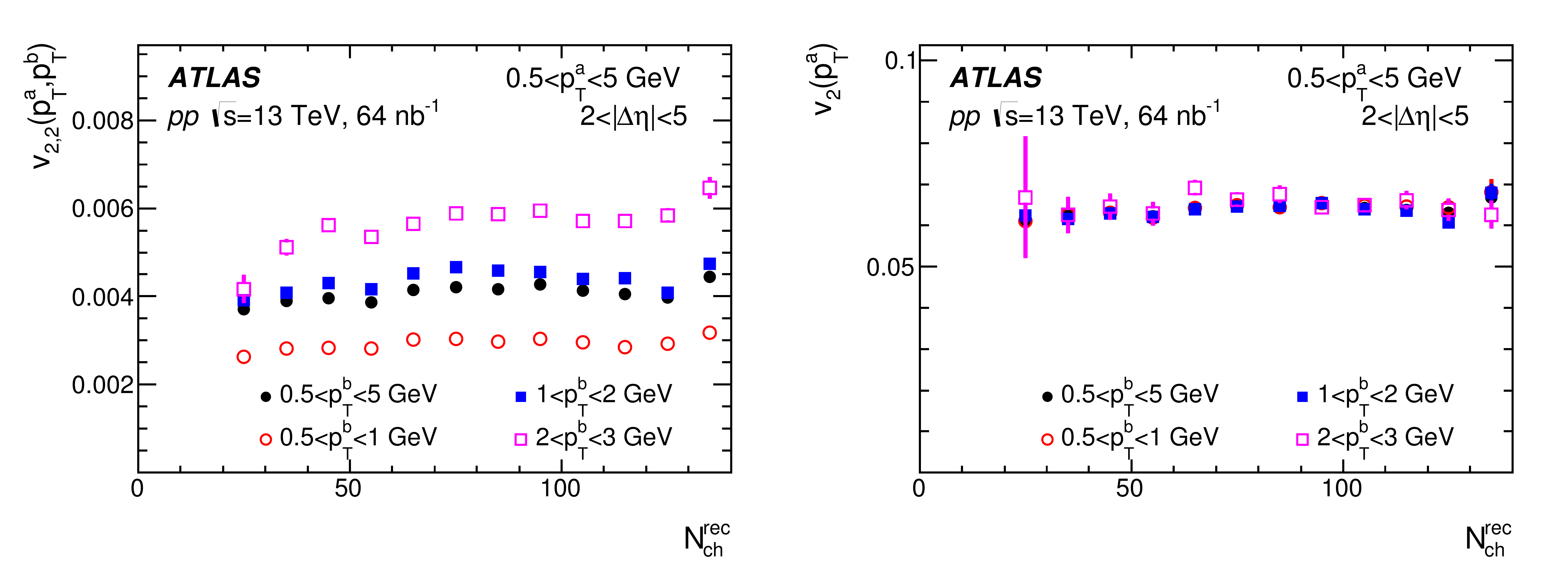}
\end{centering}
\vspace{-0.5cm}
\caption{
The left panel shows \vtt as a function of \nchrec in the 13~\TeV\ \pp data,
  for $0.5{<}$\pta${<}5$~\GeV\ and for different choices of the 
  \ptb interval.
The right panel shows the corresponding \vtwo values obtained using Eq.\,\eqref{eq:factortwo}.
The error bars indicate statistical uncertainties only. 
\label{fig:pp_factorize_centdep}
}
\end{figure}

\begin{figure}
\begin{centering}
\includegraphics[width=1.0\linewidth]{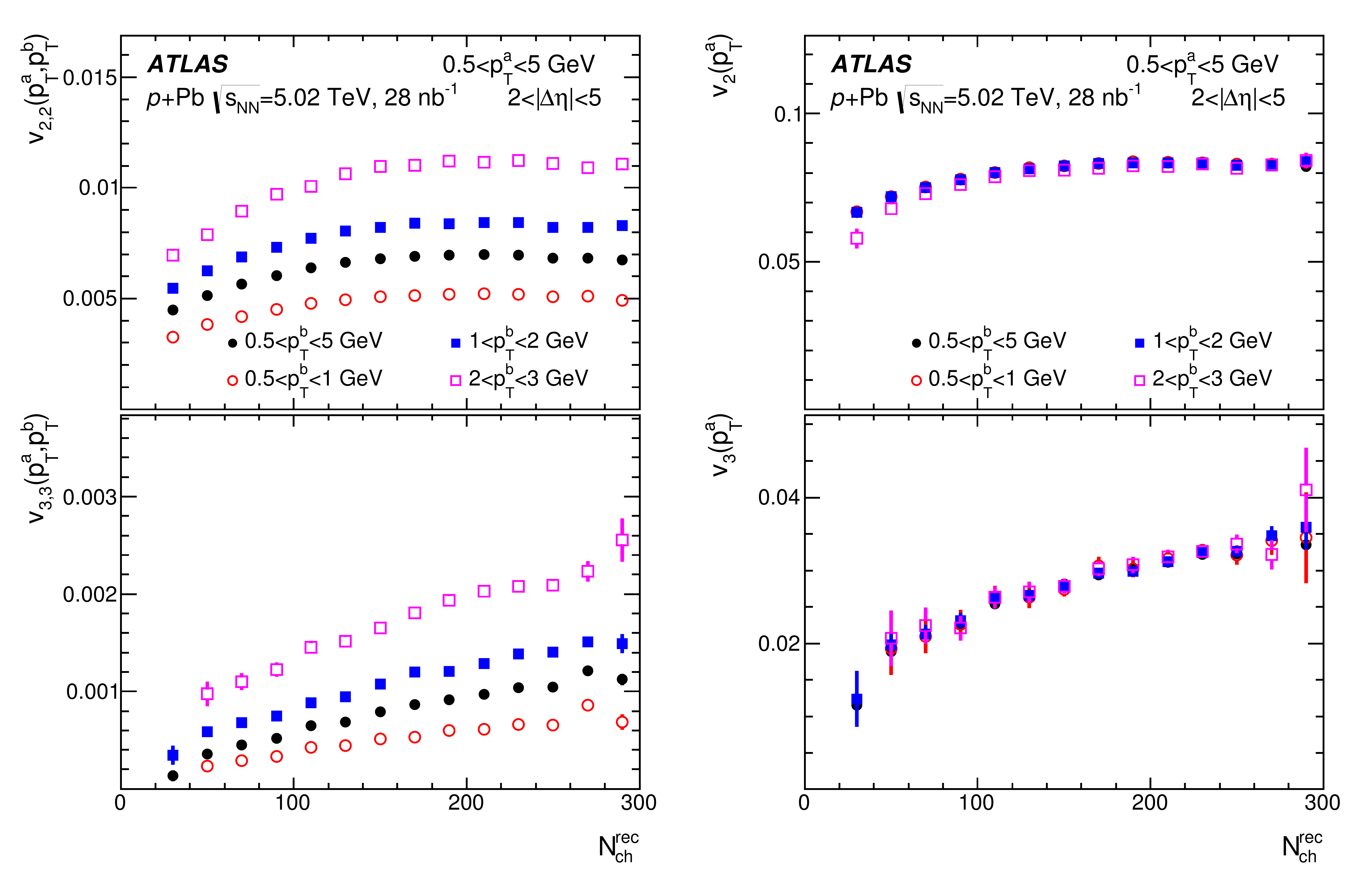}
\end{centering}
\vspace{-0.5cm}
\caption{
The left panels show \vtt (top) and \vthth (bottom) as a function of 
  \nchrec in the 5.02~\TeV\ \pPb data, for $0.5{<}\pta{<}5$~\GeV\ and for 
  different choices of the  \ptb interval.
The right panels shows the corresponding \vtwo (top) and \vthree (bottom) 
  values obtained using Eq.\,\eqref{eq:factortwo}.
The error bars indicate statistical uncertainties only. 
\label{fig:pPb_factorize_centdep}
}
\end{figure}

If the \vnn obtained from the template fits are the result of single-particle
  modulations, then the \vnn should factorize 
  as in Eq.\,\eqref{eq:factorone}, and the $\vn(\pta)$ obtained by
  correlating trigger particles at a given \pta with associated
  particles in several different intervals of \ptb (Eq.\,\eqref{eq:factortwo})
  should be independent of the choice of the \ptb interval.
Figure~\ref{fig:pp_factorize_centdep} demonstrates the factorization 
  of the \vtt in the 13~\TeV\ \pp data, as a function of \nchrec.
The left panel shows the \vtt values for $0.5{<}\pta{<}5$~\GeV\ and for four different
  choices of the associated particle \pt: 0.5--5, 0.5--1, 1--2
  and 2--3~\GeV.
The right panel shows the corresponding $\vtwo(\pta)$ obtained using 
  Eq.\,\eqref{eq:factortwo}.
While the $\vtt(\pta,\ptb)$ values vary by a factor of $\sim$2 between
  the different choices of the \ptb interval, the corresponding
  $\vtwo(\pta)$ values agree quite well. 
Similar plots for the \pPb data are shown in Figure\,\ref{fig:pPb_factorize_centdep}.
Here due to higher statistical precision in the data,  the factorization 
  is tested for both \vtt and \vthth.
The variation of $\vtt(\pta,\ptb)$ between the four \ptb intervals
  is a factor of ${\sim} 2$ while the variation of $\vthth(\pta,\ptb)$
  is more than a factor of 3.
However, the corresponding $\vn(\pta)$ values are in good agreement
  with each other, with the only exception being the \vtt values for 
  $2{<}\ptb{<} 3$~\GeV\ where some deviation from this behavior is 
  seen for $\nchrec{\lesssim} 60$.

\begin{figure}
\begin{centering}
\includegraphics[width=1.0\linewidth]{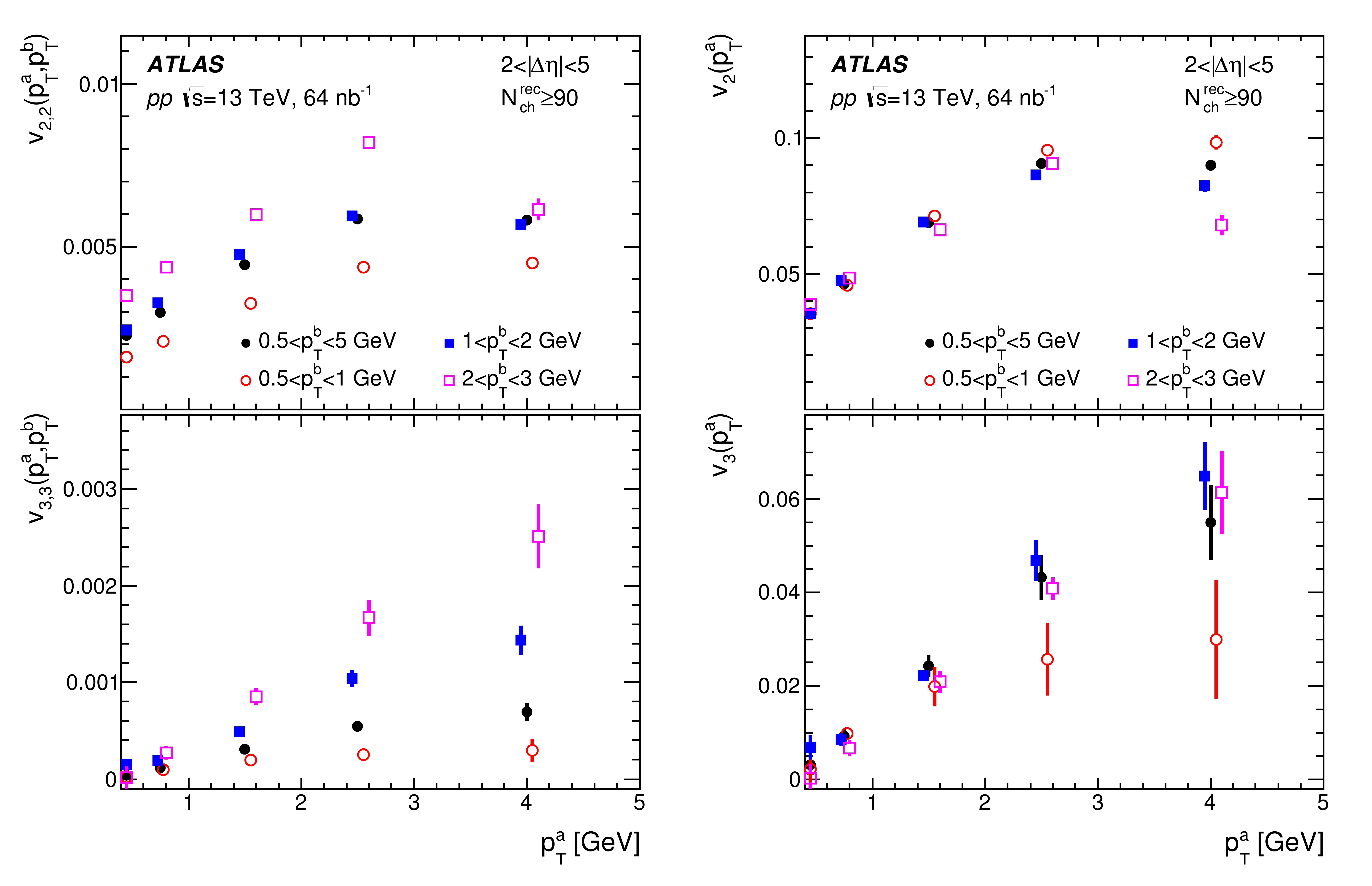}
\end{centering}
\vspace{-0.5cm}
\caption{
The left panels show \vtt (top) and \vthth (bottom) as a function of 
  \pta in the 13~\TeV\ \pp data, for \nchrec$\geq$90 and for 
  different choices of the \ptb interval.
The right panels shows the corresponding \vtwo (top) and \vthree (bottom) 
  values obtained using Eq.\,\eqref{eq:factortwo}.
The error bars indicate statistical uncertainties only. 
The \pta intervals plotted are 0.4--0.5, 0.5--1, 1--2, 2--3 and 3--5~\GeV.
In some cases, the data points have been slightly shifted along the $x$-axis, for clarity. 
\label{fig:pp_factorize_ptadep}
}
\end{figure}

\begin{figure}
\begin{centering}
\includegraphics[width=1.0\linewidth]{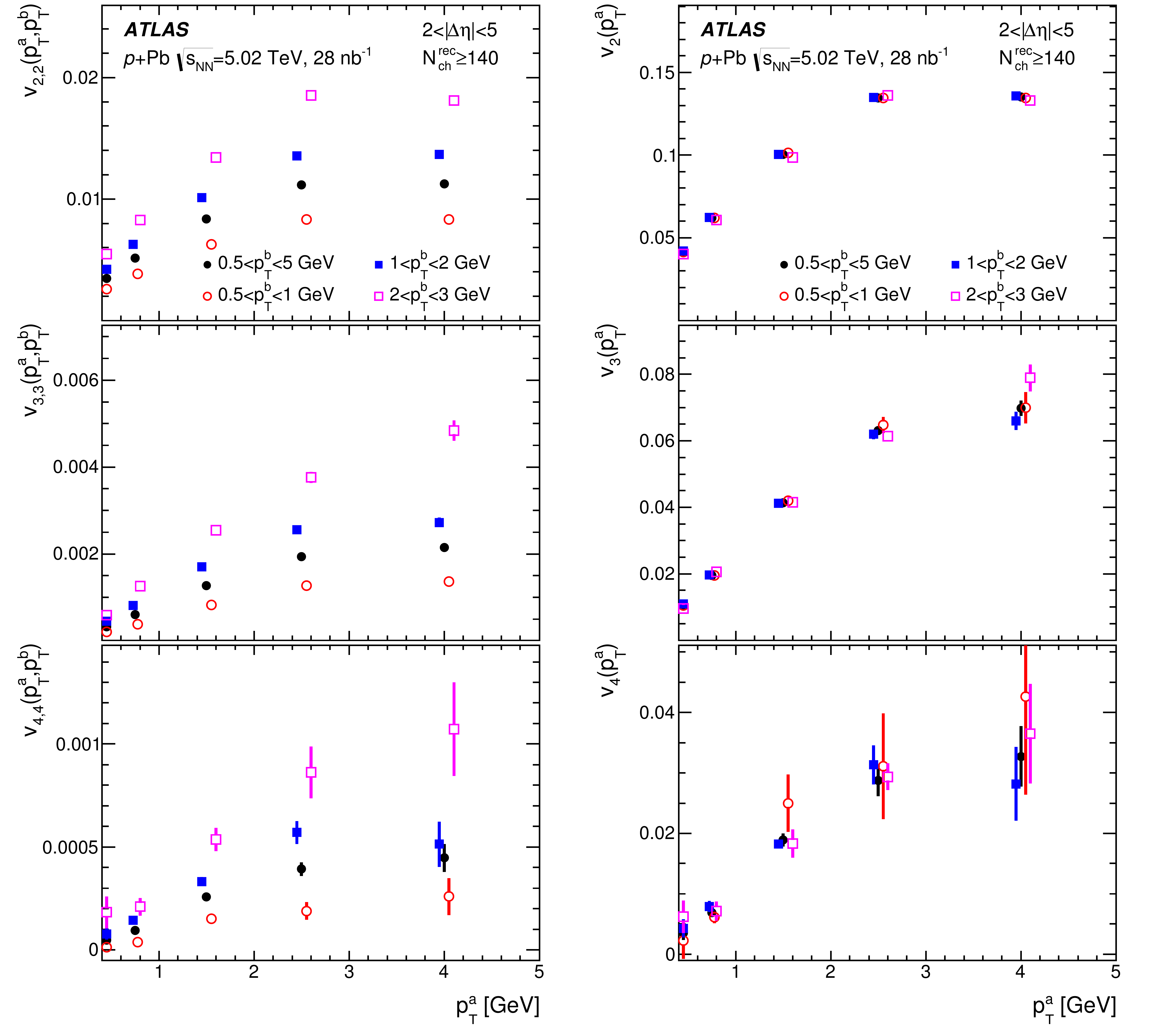}
\end{centering}
\vspace{-0.5cm}
\caption{
The left panels show the \vnn as a function of  \pta in the 5.02~\TeV\ 
  \pPb data, for \nchrec$\geq$140 and for different choices 
  of the \ptb interval.
From top to bottom, the three rows correspond to $n{=}2, 3$ and 4.
The right panels shows the corresponding \vn values obtained using 
  Eq.\,\eqref{eq:factortwo}.
The error bars indicate statistical uncertainties only. 
The \pta intervals plotted are 0.4--0.5, 0.5--1, 1--2, 2--3 and 3--5~\GeV.
In some cases, the data points have been slightly shifted along the $x$-axis, for clarity. 
\label{fig:pPb_factorize_ptadep}
}
\end{figure}

Figure~\ref{fig:pp_factorize_ptadep} studies the \pta dependence of
  the factorization in the 13~\TeV\ \pp data for \vtt (top panels)
  and \vthth (bottom panels).
The results are shown for the $\nchrec{\geq}90$ multiplicity range.
The left panels show the \vnn as a function of \pta for four different
  choices of the associated particle \pt: 0.5--5, 0.5--1, 1--2
  and 2--3~\GeV.
The right panels show the corresponding \vn(\pta) obtained 
  using Eq.\,\eqref{eq:factortwo}.
In the \vtt case, factorization holds reasonably well for $\pta{\leq} 3$~\GeV,
  and becomes worse at higher \pt. This breakdown at higher \pt\ is
  likely caused by the above-discussed multiplicity-dependent
  distortions of the dijet component of the correlation function which
  are not accounted for in the template fitting procedure.
For \vthth, the factorization holds reasonably well for $\ptb{>}1$~\GeV.
The $0.5{<}\ptb{<}1$~\GeV\ case shows a larger deviation in the factorization, 
  but has much larger associated statistical uncertainties.
Similar plots for the \pPb case are shown in Figure\,\ref{fig:pPb_factorize_ptadep}.
Here the factorization holds for \vtt, \vthth and \vff
  up to $\ptb{\sim}5$~\GeV.

\subsection{Dependence of \vnn on \deta gap}
\begin{figure}[h]
\begin{centering}
\includegraphics[width=1.0\linewidth]{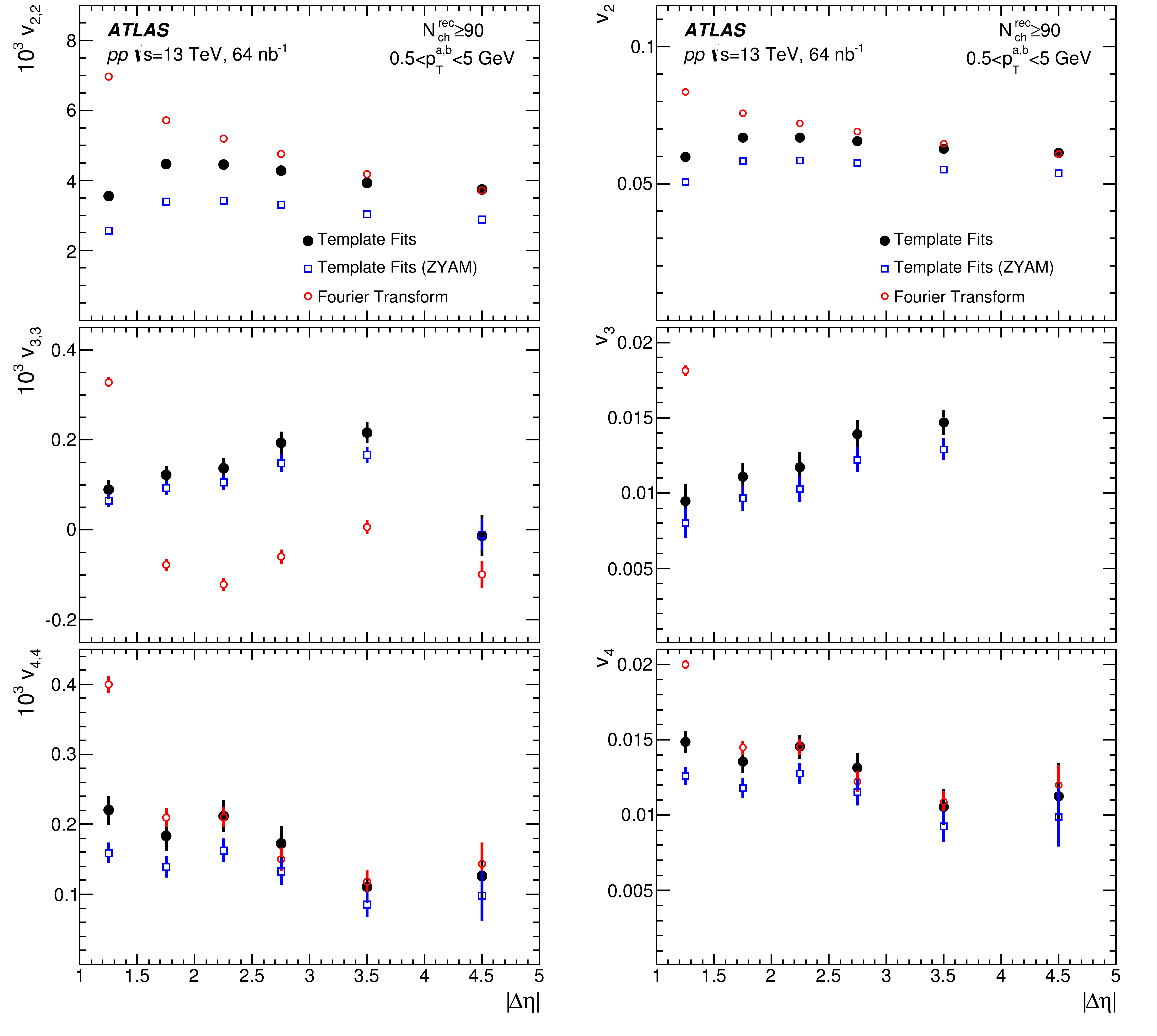}
\end{centering}
\vspace{-0.5cm}
\caption{
The $|\deta|$ dependence of the \vnn
  (left panels) and \vn (right panels) in the 13~\TeV\ \pp data. 
From top to bottom the rows correspond to $n$=2, 3 and 4, respectively.
The ZYAM-template and Fourier-\vnn values are also shown for comparison.
Only the range $|\deta|{>}1$ is shown to suppress the large Fourier-\vnn 
  at $|\deta|{\sim}0$ that arise due to the near-side jet peak.
Plots are for the $\nchrec{\geq}90$ multiplicity range and for 
  $0.5{<}\ptab{<}5$~\GeV.
The error bars indicate statistical uncertainties only.
For points where \vthth is negative,
  \vthree is not defined and hence not plotted.
\label{fig:13Tev_vnn_deta_dependence} }
\end{figure}

A systematic study of the \deta dependence of the \vnn can also help
  in determining the origin of the long-range correlation.
If it arises from mechanisms that only correlate a few particles 
  in an event, such as jets, then a strong dependence of the correlation 
  on the \deta gap between particle pairs is expected.
Figure\,\ref{fig:13Tev_vnn_deta_dependence} shows the measured \vnn
  (left panels) and \vn=$\sqrt{\vnn}$ (right panels),
  as a function of $|\deta|$ for $|\deta|$>1 in the 13~\TeV\ \pp data. 
Also shown for comparison are the Fourier and ZYAM-based template-\vnn.
The template-\vtt (top left panel) and \vtwo (top right panel) are quite stable, 
  especially for $|\deta|$>1.5, where the influence of the near-side jet is diminished.
In contrast, the Fourier-\vtt show a strong $|\deta|$ dependence.
The \deta dependence is largest at small $|\deta|$ because of the presence of the sharply
  peaked near-side jet, but is considerable even for $|\deta|$>2.
Similarly, the Fourier-\vthth shows large $|\deta|$ dependence, going from 
  positive values at $|\deta|{\sim}1$ to negative values at large $|\deta|$,
  while the template-\vthth change only weakly in comparison.
The Fourier-\vthth is often negative,
  ruling out the possibility of it being generated by single-particle
  anisotropies, which require that $v_{n,n}=v_{n}^2$ be positive.
For points where \vthth is negative,
  \vthree is not defined and hence not plotted.
The template-\vthth is, however, positive and, therefore, consistent with a 
  single-particle anisotropy as its origin, except for the highest
  $|\deta|$ interval where it is consistent with zero.
The \vff values, like the \vtt and \vthth values, vary only weakly with $|\deta|$.
These observations further support the conclusion that the template-\vnn 
  are coefficients of genuine long-range correlations.

\FloatBarrier
\section{Systematic uncertainties and cross-checks}
\label{sec:systematics}
The systematic uncertainties in this analysis arise from
  choosing the peripheral bin used in the template fits, 
  pileup,
  tracking efficiency, 
  pair-acceptance 
  and 
  Monte Carlo consistency.
Each source is discussed separately below.

\textbf{Peripheral interval:} 
As explained in Section\,\ref{sec:fits}, the template fitting procedure makes
  two assumptions. 
First it assumes that the contributions to \Yphi from hard processes 
  have identical shape across all event activity ranges,
  and only change in overall scale.
Second, it assumes that the \vnn are only weakly dependent on the 
  event activity.
The assumptions are self-consistent for the \nchrec dependence of the
  \vnn in the 5.02 and 13~\TeV\ \pp data (Figures\,\ref{fig:fit_parms_13Tev}--\ref{fig:fit_parms_5Tev}),
  where the measured template-\vnn values do turn out to be nearly independent
  of \nchrec.
However, for the \fcalet dependence in the \pp data, and for both the 
  \nchrec and \fcaletPb dependence in the \pPb data,  a systematic 
  increase of the template-\vtt with event activity is seen at small event activity.
This indicates the breakdown of one of the above two assumptions.
To test the sensitivity of the measured \vnn to any residual changes
  in the width of the away-side jet peak and to the \vnn present in
  the peripheral reference, the analysis is repeated using 0${\leq}\nchrec{<}$10 and 
  10${\leq}\nchrec{<}$20 intervals to form \YphiPer.
The variations in the \vnn for the different chosen peripheral 
  intervals are taken to be a systematic uncertainty.
For a given dataset, this uncertainty is strongly correlated 
  across all multiplicity intervals.
Choosing a peripheral interval with larger mean multiplicity typically decreases
  the measured \vnn.

\begin{figure}
\begin{centering}
\includegraphics[width=1.0\linewidth]{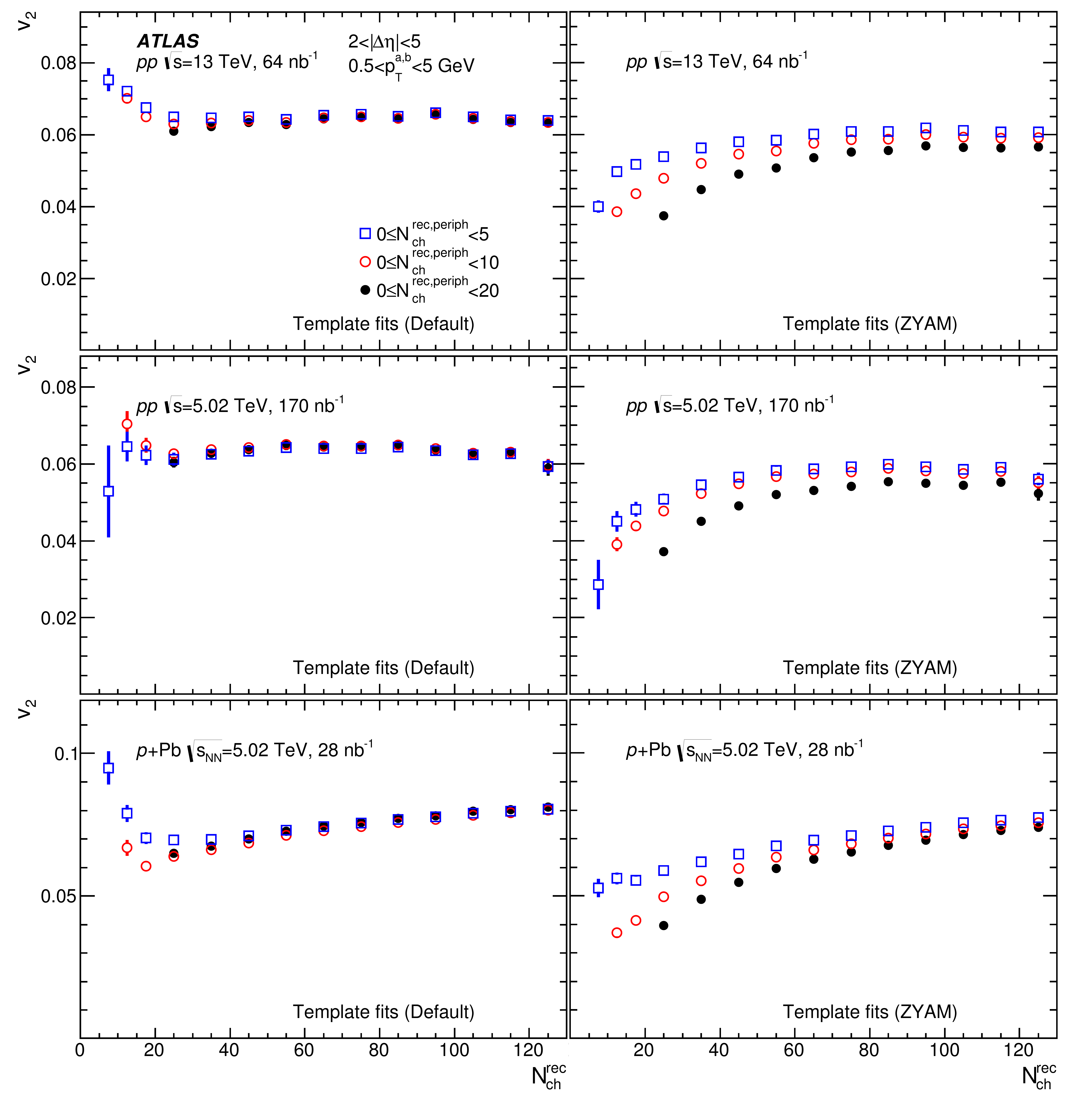}
\end{centering}
\caption{
Dependence of \vtwo on the peripheral bin chosen for
  the default (left panels) and ZYAM-based (right panels) template fitting
  methods. 
The top panels correspond to 13~\TeV\ \pp collisions, middle
  panels correspond to 5.02~\TeV\ \pp collisions, and the lower panels
  correspond to 5.02~\TeV\ \pPb collisions. 
The results are plotted as a function of \nchrec and for 0.5<\ptab<5~\GeV.
The error bars indicate statistical uncertainties.
\label{fig:v22_peripheral_bin_dep}
}
\end{figure}

The sensitivity of the template-\vtwo to which 
   peripheral interval is chosen is demonstrated in the left panels of 
   Figure\,\ref{fig:v22_peripheral_bin_dep}, where $v_2$ is shown
   for three different peripheral \nchrec interval choices: 0${\leq}\Ntrkperi{<}$5, 0${\leq}\Ntrkperi{<}$10 and 0${\leq}\Ntrkperi{<}$20.
In both the 13~\TeV\ and 5.02~\TeV\ \pp data, except at very low \nchrec, 
   the \vtwo values are nearly independent of the chosen peripheral reference.
In the 13~\TeV\ \pp case, the variation is ${\sim} 6\%$ at 
   $\nchrec{\sim}30$ and decreases to ${\sim}1\%$ for $\nchrec{\geq}60$.
Even in the \pPb case, where the measured template-\vtt exhibits some 
   dependence on \nchrec, the dependence of the template-\vtwo on the 
   choice of peripheral bin is quite small: ${\sim}6\%$ at $\nchrec{\sim}30$
   and decreases to ${\sim}2\%$ for $\nchrec{\sim}60$.
Also shown for comparison are the corresponding \vtwo values obtained 
   from the ZYAM-based template fitting method 
   (right panels of Figure\,\ref{fig:v22_peripheral_bin_dep}).
These exhibit considerable dependence on the peripheral reference.
For the 13~\TeV\ \pp case, the variation in the ZYAM-based \vtwo is
  ${\sim}40\%$ at $\nchrec{\sim}30$, and decreases to ${\sim}12\%$ at $\nchrec{\sim}$60 
  and asymptotically at large \nchrec is ${\sim} 7\%$.
For the \pPb case, the variation is even larger:
   ${\sim}35\%$ at $\nchrec{\sim}30$ and ${\sim}14\%$ for $\nchrec\sim$60. 
These results show that the template-\vtwo is quite stable 
   as the peripheral interval is varied, while the ZYAM-based result is very sensitive.
This is one of the advantages of the new method.
For the ZYAM-based results, as the upper edge of the peripheral interval is 
  moved to lower multiplicities, the measured \vtwo becomes 
  less and less dependent on \Ntrk.
Qualitatively, it seems that in the limit of $\Ntrkperi\rightarrow 0$ the ZYAM-based \pp-\vtwo 
  would be nearly independent of \Ntrk, thus contradicting the assumption of zero \vtwo
  made in the ZYAM method, and supporting the flat-\vtwo assumption made in the new method.

\textbf{Pileup:}
Pileup events, when included in the two-particle correlation measurement,
  dilute the \vnn signal since they produce pairs where the trigger and associated 
  particle are from different collisions and thus have no physical correlations.
The maximal fractional dilution in the \vnn is equal to the pileup rate.
In the \pPb data, nearly all of the events containing pileup are removed 
  by the procedure described in Section\,\ref{sec:data}. 
The influence of the residual pileup is evaluated by relaxing the pileup 
  rejection criteria and then calculating the change in the \Yphi and
  \vn values. 
The differences are taken as an estimate of the uncertainty for the \vnn, 
  and are found to be negligible in low event activity classes, and increase to 4\% 
  for events with $\nchrec{\sim}300$.

In the \pp data, for events containing multiple vertices, only tracks 
  associated with the vertex having the largest $\sum\pt^2$, where the sum
  is over all tracks associated with the vertex, are used in the analysis.
Events with multiple unresolved vertices
  affect the results by increasing the combinatoric pedestal in \Yphi.
The fraction of events with merged vertices is estimated and taken as
  the relative uncertainty associated   
  with pileup in the \pp analysis.
The merged-vertex rate in the 13~\TeV\ \pp data is 0--3\% over the 
  0--150 \nchrec range.
In the 5.02 \TeV\ \pp\ data, it is 0--4\% over the 0--120 \nchrec range.

\textbf{Track reconstruction efficiency:} 
In evaluating \Yphi, each particle is weighted by 
  $1/\epsilon(\pt,\eta)$ to account for the tracking efficiency. 
The systematic uncertainties in the efficiency $\epsilon(\pt,\eta)$ thus need 
  to be propagated into \Yphi and the final \vnn measurements.
Unlike \Yphi, which is strongly affected by the efficiency, 
  the \vnn are mostly insensitive to the tracking efficiency. 
This is because the \vnn measure the relative variation of the yields 
  in \dphi; an overall increase or decrease in the efficiency changes the 
  yields but does not affect the \vnn.
However, as the tracking efficiency and its uncertainties have \pt and
  \eta dependence, there is some residual effect on the \vnn.
The corresponding uncertainty in the \vnn is estimated by repeating the 
  analysis while varying the efficiency to its upper and lower extremes.
In the \pp analysis, this uncertainty is estimated to be 0.5\% for \vtt 
  and 2.5\% for \vthth and \vff.
The corresponding uncertainties in the $\pPb$ data are 0.8\%, 1.6\% and 2.4\%
  for \vtt, \vthth and \vff, respectively.

\textbf{Pair-acceptance:} 
As described in Section\,\ref{sec:corr}, this analysis uses the mixed-event 
  distributions $B(\deta,\dphi)$ and $B(\dphi)$ to estimate and correct 
  for the pair-acceptance of the detector.
The mixed-event distributions are in general quite flat in $\dphi$.
The Fourier coefficients of the mixed-event distributions, $\vnn^{\mathrm{det}}$,
  which quantify the magnitude of the corrections, are ${\sim} 10^{-4}$
  in the \pPb data, and ${\sim} 2\times10^{-5}$ in the \pp data.
In the \pPb analysis, potential systematic uncertainties in the \vnn 
  due to residual pair-acceptance effects not corrected by the 
  mixed-events are evaluated following Ref.\,\cite{HION-2011-01}.
This uncertainty is found to be smaller than ${\sim}10^{-5}$.
In the \pp analysis, since the mixed-event corrections are themselves quite small,
  the entire correction is conservatively taken as the systematic uncertainty.

\textbf{MC closure:} 
The analysis procedure is validated by measuring the \vnn of
  reconstructed particles in fully simulated \PYEight and HIJING events and
  comparing them to those obtained using the generated particles.
The difference between the generated and reconstructed \vnn varies between 
  $10^{-5}$ and $10^{-4}$ (absolute) in the \pp case and between 2\% and 8\% (relative) 
  in the \pPb case, for the different harmonics.
This difference is an estimate of possible systematic effects
  that are not accounted for in the measurement,
  such as a mismatch between the true and reconstructed momentum for charged particles,
  and is included as a systematic uncertainty.

\begin{figure}[h]
\centering
\includegraphics[width=1.0\linewidth]{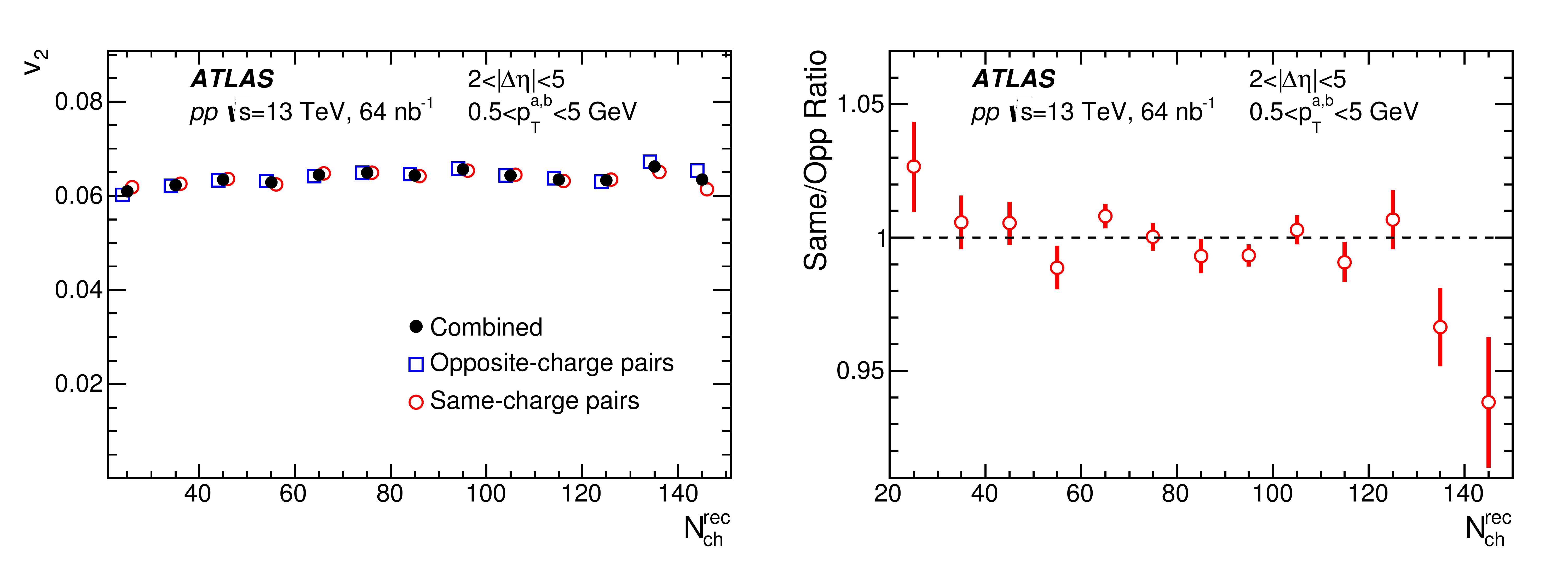}
\caption{
Left panel: comparison of the \vtwo for same-charge and opposite-charge 
  pairs in the 13~\TeV\ \pp data.
Also shown are the \vtwo values for the two types of pairs combined.
Right panel: ratio of the same-charge to the opposite-charge \vtwo.
The results are for 0.5<\ptab<5~\GeV. 
For clarity, the data points for the same-charge and opposite-charge \vn are 
  slightly shifted along the $x$-axis,.
The error bars indicate statistical uncertainties.
\label{fig:13Tev_vnn_charge_dependence}}
\end{figure}

As a cross-check, the dependence of the long-range correlations on the relative charge of the
  two particles used in the correlation is studied.
If the long-range correlations arise from phenomena that correlate only a 
  few particles in an event, such as jets or decays, 
  then a dependence of the correlation on the relative sign 
  of the particles making up the pair is expected.
Figure\,\ref{fig:13Tev_vnn_charge_dependence} shows the measured \vtwo
  from the template fits for both the same-charge and opposite-charge pairs. 
No systematic difference between the two is observed.

Tables\,\ref{tab:List-of-systematic-1} and~\ref{tab:List-of-systematic-1a} 
  list the systematic uncertainties in the \vnn for the 13~\TeV\ and 5.02~\TeV\ \pp 
  data, respectively.
Most uncertainties are listed as relative uncertainties (in percentages of the \vnn), 
  while some are listed as absolute uncertainties.
Uncertainties for the \pPb data are listed in Table\,\ref{tab:List-of-systematic-2}.
The corresponding uncertainties in the \vn are obtained by propagating the uncertainties 
  in the \vnn when using Eq.\,\eqref{eq:factorone} to obtain the \vn.
In some cases the systematic uncertainties in the \vnn are larger than 100\%. 
In these cases the corresponding uncertainties in the \vn cannot be 
calculated, as the \vn are only defined for $\vnn{>}0$. 
Such cases are excluded from the \vn results presented in Section\,\ref{sec:results} below.

\begin{table}[h]
\begin{centering}
\begin{tabular}{|c| cc| cc| cc|}
\hline 
  \multirow{3}{*}{Source}     & \multicolumn{2}{c|}{\vtt}
                              & \multicolumn{2}{c|}{\vthth}
                              & \multicolumn{2}{c|}{\vff} \tabularnewline \cline{2-7}
                              & \nchrec  & syst.         & \nchrec & syst.  & \nchrec  & syst.  \tabularnewline 
                              &  range   & unc.      &  range & unc.      & range & unc.      \tabularnewline \hline \hline

Choice of peripheral bin [\%] & 20--30 & 7              & 20--50  & ${>}$100 & 20--50           &30     \tabularnewline 
 0.5<\ptab<5~\GeV\            & 30--60 & 7--2           & 50--100 & 100--40  & 50--100          &30--10 \tabularnewline 
                              & ${>}60$& 2              & ${>}100$& 40       & ${>} 100$        &10     \tabularnewline\hline
Choice of peripheral bin [\%] & 20--30 & 6              & 20--60  & 40--20   &                  &       \tabularnewline 
 1<\ptab<5~\GeV\              & 30--60 & 6--2           & 60--100 & 20--10   &                  & 5     \tabularnewline 
                              & ${>}80$& 2              & ${>}100$& 10       &                  &       \tabularnewline\hline
Pileup  [\%]                  & 0--150 & 0--3           &  0--150 & 0--3     &    0--150        &0--3   \tabularnewline\hline 
Tracking efficiency [\%]      &        & 0.5            &         & 2.5      &                  &2.5    \tabularnewline\hline 
Pair acceptance (absolute)    &        &$2\times10^{-5}$&&  $2\times10^{-5}$ && $2\times10^{-5}$        \tabularnewline\hline 
MC closure (absolute)         &        &$1\times10^{-4}$&&  $2\times10^{-5}$ && $2\times10^{-5}$        \tabularnewline\hline
\end{tabular}
\par\end{centering}
\caption{
Systematic uncertainties for the \vnn obtained from the
  template analysis in the 13~\TeV\ \pp data.
Where ranges are provided for both multiplicity and the uncertainty, 
   the uncertainty varies from the first value to the second value 
   as the multiplicity varies from the lower to upper limits of the range. 
Where no multiplicity range is provided the uncertainty is 
   multiplicity-independent.
\label{tab:List-of-systematic-1}}
\end{table}

\begin{table}[h]
\begin{centering}
\begin{tabular}{|c| cc| cc| cc|}
\hline 
  \multirow{3}{*}{Source}     & \multicolumn{2}{c|}{\vtt}
                              & \multicolumn{2}{c|}{\vthth}
                              & \multicolumn{2}{c|}{\vff} \tabularnewline \cline{2-7}
                              & \nchrec  & syst.         & \nchrec & syst.  & \nchrec  & syst.  \tabularnewline 
                              &  range   & unc.      &  range & unc.      & range & unc.      \tabularnewline \hline \hline

Choice of peripheral bin [\%] & 20--30 & 8              & 20--30  & 55       &   20--30  &${>}100$\tabularnewline 
 1<\ptab<5~\GeV\                & 30--70 & 8--2           & 30--50  & 55--12   &  ${>}30$  & 50     \tabularnewline 
                              & ${>}70$& 2              & ${>}50$ & 12       &           &        \tabularnewline\hline
Pileup  [\%]                  & 0--120 & 0--4           & 0--120  & 0--4     & 0--120    &0--4    \tabularnewline\hline 
Tracking efficiency [\%]      &        & 0.5            &         & 2.5      &                  &2.5    \tabularnewline\hline 
Pair acceptance (absolute)    &        &$2\times10^{-5}$&&  $2\times10^{-5}$ && $2\times10^{-5}$        \tabularnewline\hline 
MC closure (absolute)         &        &$1\times10^{-4}$&&  $2\times10^{-5}$ && $2\times10^{-5}$        \tabularnewline\hline
\end{tabular}
\par\end{centering}
\caption{
Systematic uncertainties for the \vnn obtained from the
  template analysis in the 5.02~\TeV\ \pp data.
Where ranges are provided for both multiplicity and the uncertainty, 
   the uncertainty varies from the first value to the second value 
   as the multiplicity varies from the lower to upper limits of the range. 
Where no multiplicity range is provided the uncertainty is 
   multiplicity-independent.
\label{tab:List-of-systematic-1a}}
\end{table}

\begin{table}[h]
\begin{centering}
\begin{tabular}{|c| cc| cc| cc|}
\hline 
  \multirow{3}{*}{Source}     & \multicolumn{2}{c|}{\vtt}
                              & \multicolumn{2}{c|}{\vthth}
                              & \multicolumn{2}{c|}{\vff} \tabularnewline \cline{2-7}
                              & \nchrec  & syst.     & \nchrec  & syst.    & \nchrec   & syst.      \tabularnewline 
                              &  range   & unc.      &  range   & unc.     & range     & unc.       \tabularnewline \hline \hline
Choice of peripheral bin [\%] &  20--30  & 5         &  20--30  & ${>}100$ &  20--30   & ${>} 100$  \tabularnewline 
0.5<\ptab<5~\GeV\               &  30--250 & 5--2      &  30--50  & 100--40  &  30--50   & 100--20    \tabularnewline 
                              &          &           & 50--250  & 40--5    &  50--250  &  20--2     \tabularnewline\hline
Choice of peripheral bin [\%] &  20--30  & 12        &  20--50  & 55--20   &  20--50   &  70--10    \tabularnewline 
1<\ptab<5~\GeV\               &  30--50  & 12--6     &  50--100 & 20--10   &  50--250  &  10--5     \tabularnewline 
                              &  50--250 & 6--2      & 100--250 & 10--5    &           &            \tabularnewline\hline
 Pileup [\%]                  &   0--300 & 0--4      &  0--300  &0--4      &   0--300  &0--4        \tabularnewline\hline 
 Tracking efficiency [\%]     &          & 0.8       &          &1.6       &           &2.4         \tabularnewline\hline 
 Pair acceptance (absolute)   &          & $10^{-5}$ &          &$10^{-5}$ &           &$10^{-5}$   \tabularnewline\hline 
 MC closure [\%]              &          & 2         &          &4         &           &8           \tabularnewline\hline
\end{tabular}
\par\end{centering}
\caption{
Systematic uncertainties for the \vnn obtained from the
   template analysis in the 5.02~\TeV\ \pPb data. 
Where ranges are provided for both multiplicity and the uncertainty, 
   the uncertainty varies from the first value to the second value 
   as the multiplicity varies from the lower to upper limits of the range. 
Where no multiplicity range is provided the
   uncertainty is multiplicity-independent.
\label{tab:List-of-systematic-2}}
\end{table}

\FloatBarrier
\section{Results}
\label{sec:results}
\begin{figure}
\begin{centering}
\includegraphics[width=1.0\linewidth]{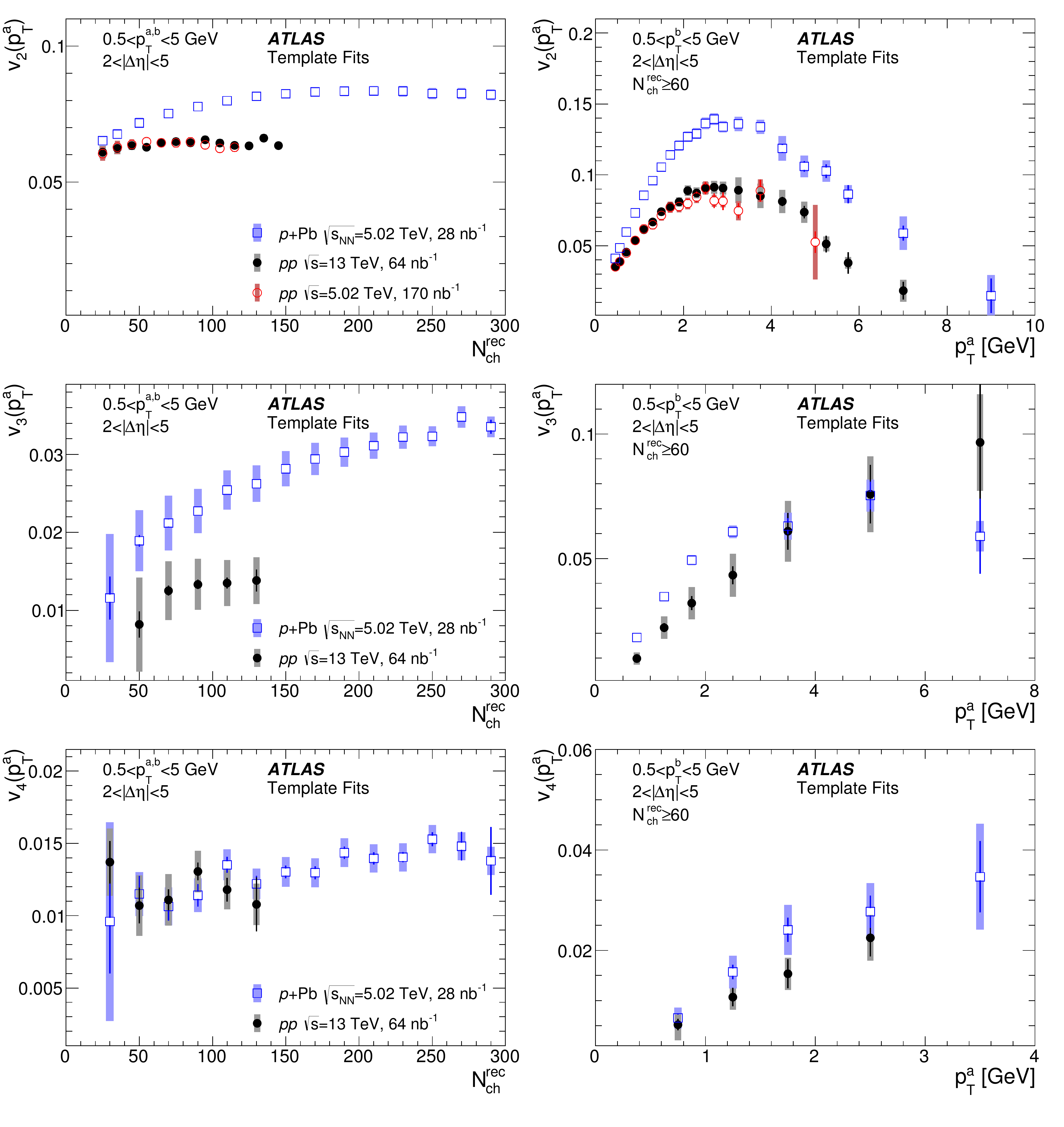}
\end{centering}
\caption{
Left panels: comparison of the \vn obtained from the template
  fitting procedure in the 13~\TeV\ \pp, 5.02~\TeV\ \pp, and 5.02~\TeV\
  \pPb data, as a function of \nchrec. 
The results are for $0.5{<} \ptab {<} 5$~\GeV. 
Right panels: the \pt dependence of the \vn for the $\nchrec{\geq}60$
  multiplicity range.
From top to bottom the rows correspond to $n$=2, 3 and 4, respectively.
The error bars and shaded bands indicate statistical and systematic uncertainties, respectively.
\label{fig::compare_systems}
}
\end{figure}

\begin{figure}
\includegraphics[width=1.0\linewidth]{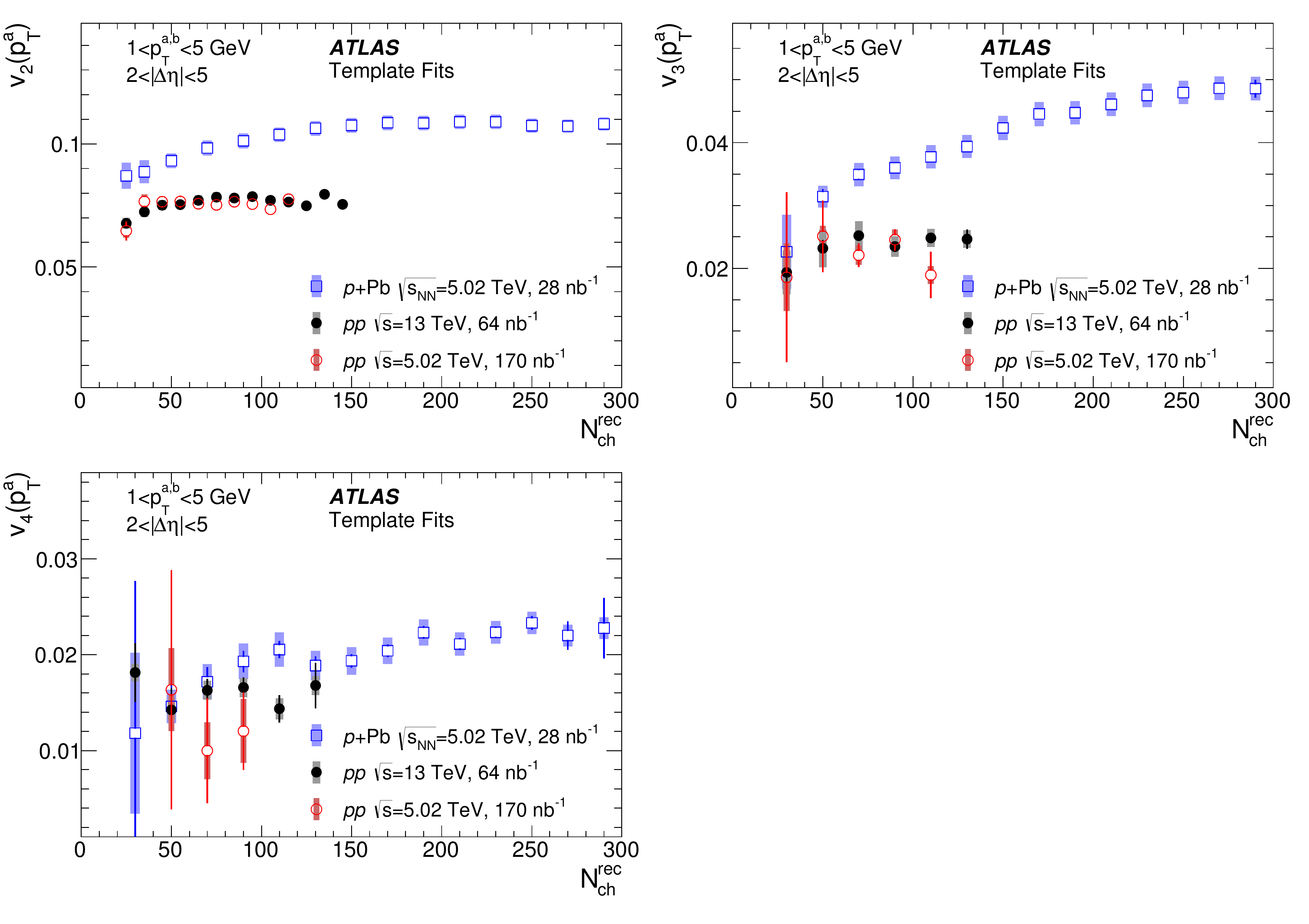}
\caption{
Comparison of the \vn obtained from the template
  fitting procedure in the 13~\TeV\ \pp, 5.02~\TeV\ \pp, and 5.02~\TeV\
  \pPb data, as a function of \nchrec. 
The results are for $1{<}\ptab{<}5$~\GeV. 
The three panels correspond to $n$=2, 3 and 4.
The error bars and shaded bands indicate statistical and systematic uncertainties, respectively.
\label{fig::compare_systems3}
}
\end{figure}

Figure\,\ref{fig::compare_systems}  provides a summary of the main
  results of this \papertype\ in the inclusive \pT\ interval $0.5 {<} \pT {<}
  5$~\GeV. 
It compares the \vn obtained from
  the 5.02~\TeV, 13~\TeV\ \pp and 5.02~\TeV\ \pPb
  template fits. 
The left panels show $v_2, v_3$ and $v_4$ as a function of 
  \nchrec\ while the right panels show the results as a function of \pta 
  for the $\nchrec{\geq}60$ multiplicity range.
The measured \vthree and \vfour in the 5.02~\TeV\ \pp data for $0.5{<}\ptab{<}5$~\GeV\ 
  have large systematic uncertainties associated with the choice of
  peripheral reference and are not shown in Figure\,\ref{fig::compare_systems}.
They are shown in Figure\,\ref{fig::compare_systems3} for a different \pT\ interval
  of $1{<}\ptab{<}5$~\GeV.
Figure\,\ref{fig::compare_systems} shows that the \pPb\ $v_2$ 
  increases with increasing \Ntrk\ as previously observed~\cite{HION-2013-04} 
  while the \pp\ $v_2$ is \Ntrk-independent within uncertainties. 
The \pPb\ $v_3$ is significantly larger than the $\pp$ \vthree and 
  also shows a systematic increase with \Ntrk, 
  while the \pp\ $v_3$ is consistent with being \Ntrk-independent. 
The \pp\ and \pPb\ $v_4$ are consistent within large uncertainties, 
  and the \pPb\ $v_4$ increases weakly with increasing \Ntrk. 

The difference between the \pp\ and \pPb\ results for the \Ntrk\
dependence of the $v_n$ is expected. Studies of the
centrality dependence of the multiplicity distributions in \pPb\
collisions show a strong correlation between the multiplicity and the
number of participants, or equivalently, the number of scatterings of
the proton in the nucleus~\cite{HION-2012-15}. Regardless of the
interpretation of the results, a dependence of the $v_n$ on the
geometry of the \pPb\ collisions is expected~\cite{Dusling:2015gta}. 
In contrast, the relationship between
multiplicity and geometry in \pp\ collisions is poorly
understood and necessarily different as there are, by definition, only
two colliding nucleons. 
However, an early study of this problem accounting for
perturbative evolution did predict a weak dependence of $v_2$
on multiplicity, as observed in this measurement~\cite{Avsar:2010rf}. 
A more recent study that models the proton substructure and fluctuations 
   in the multiplicity of the final particles, showed that
   the eccentricities $\epsilon_2$ and $\epsilon_3$ of the initial 
   entropy-density distributions in \pp\ collisions have no correlation 
   with the final particle multiplicity~\cite{Welsh:2016siu}.
If the \vn in \pp\ collisions are directly related to the $\epsilon_n$, 
   then the calculations in Ref.\,\cite{Welsh:2016siu} are consistent 
   with the trends observed in the measured \vn.

\begin{figure}[h]
\centering
\includegraphics[width=0.65\linewidth]{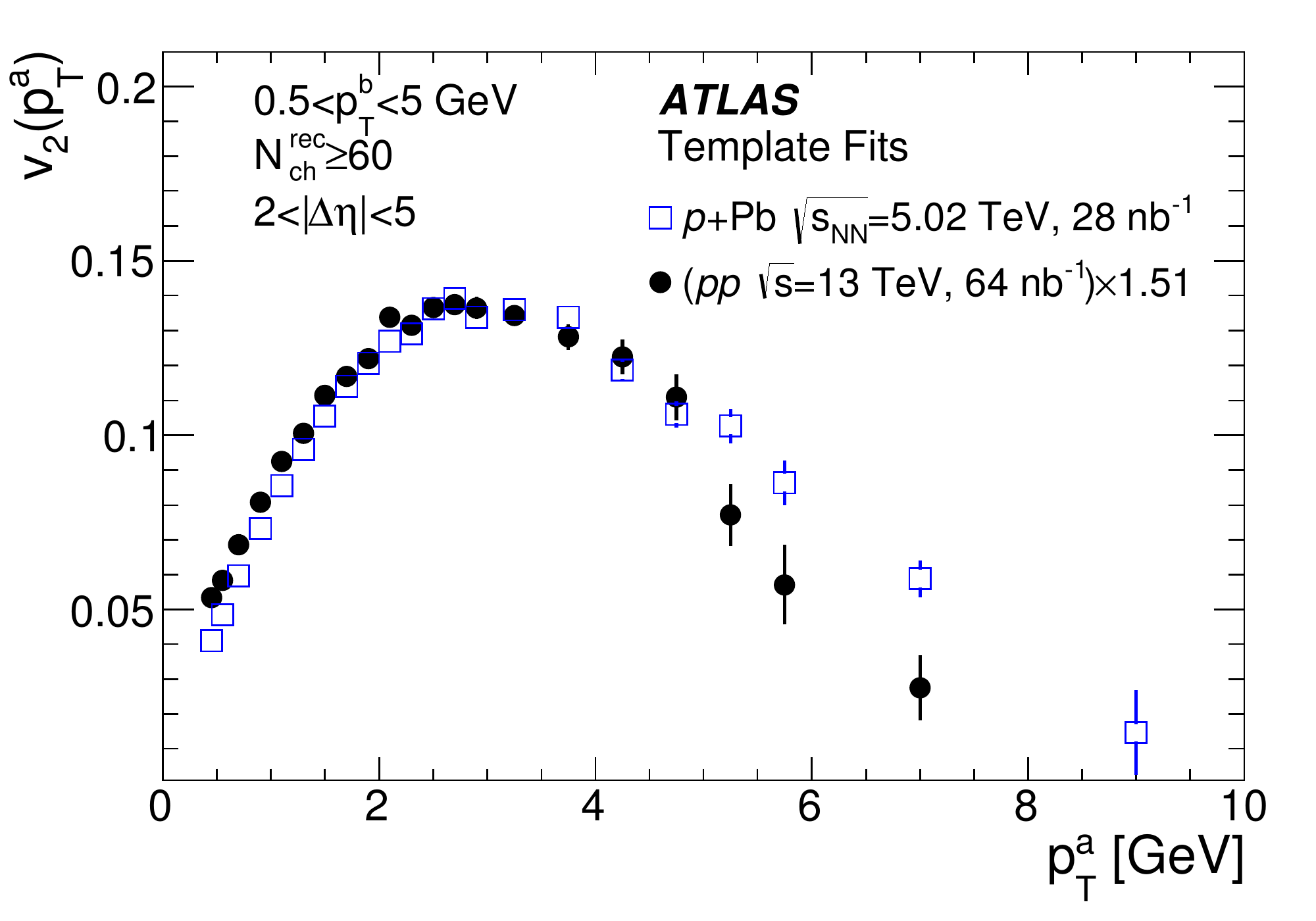}
\caption{
Comparison of the shapes of the $\vtwo(\pt)$ in the
  13~\TeV\ \pp and 5.02~\TeV\ \pPb data.
The \pp \vtwo has been scaled by a factor of 1.51 along 
  the $y$-axis in order to match the maximum of the \vtwo in
  the two data sets.
The results are for 0.5<\ptb<5~\GeV and $\nchrec\geq$60. 
The error bars indicate statistical uncertainties.
\label{fig::compare_systems_v2_scaling}
}
\end{figure}

The \pp\ and \pPb\ $v_2(\pT)$ shown in
  Figure\,\ref{fig::compare_systems} display similar trends with both
  increasing with \pT\ at low \pT, reaching a maximum near 3~\GeV\
  and decreasing at higher \pT.
The $v_2(\pT)$ values for the 5.02 and 13~\TeV\ \pp\ data agree 
  within uncertainties. 
The \pT\ dependence of the $v_3$ and $v_4$ values is similar 
  to that of $v_2$ at low \pT, where the \pPb\ results increase more rapidly with
  increasing \pT. 
However, unlike for $v_2$, the values of $v_3$ and $v_4$ are similar at high \pt\ 
  for the \pp\ and \pPb\ data.
A direct test of the similarity of the \pT\ dependence of the Fourier
coefficients in \pp\ and \pPb\ collisions is provided in
Figure\,\ref{fig::compare_systems_v2_scaling} for $n = 2$. 
The \pp\ $v_2$ values
have been multiplied by 1.51, the ratio (\pPb\ to \pp) of the
maximum $v_2$ in the top right panel in
Figure\,\ref{fig::compare_systems}. The resulting $v_2(\pta)$
values for (scaled) \pp\ and \pPb\ agree well for \pta\
up to 5~\GeV. At higher \pta\ the \pp\ $v_2$ decreases
more rapidly due to the above-described multiplicity-dependent change
in the shape of the dijet peak in the two-particle correlation
function at high \pT. After the scaling, the \pp\ $v_2(\pta)$
are slightly higher than the \pPb\ at low \pta, but the
similarity of the shapes of the \pt dependence is, nonetheless, striking. 

A separate evaluation  of the \Ntrk-dependence of the $v_2$, \vthree 
  and \vfour values  is shown in Figure\,\ref{fig::compare_systems3} 
  for the $1{<}\ptab{<}5$~\GeV\ interval, where the 5.02~\TeV\ \pp 
  measurements yield meaningful $v_3$ and $v_4$ results. 
The figure shows agreement between the 5.02 and 13~\TeV\ \pp\ data for 
  all three Fourier coefficients. 
It also shows that the \pPb\ $v_2, v_3$ and $v_4$ rise 
  monotonically with increasing \Ntrk\ while the \pp results are 
  generally \Ntrk-independent. 
One possible exception to this statement is that the 13~\TeV\ data 
  indicate a small (${\sim} 15\%$) decrease in $v_2$ in the two lowest
  \Ntrk\ intervals. 
The \pp\ and \pPb\ $v_3$ and $v_4$ agree at low \Ntrk\ while $v_2$
  still differs significantly, although by a smaller amount than
  at larger \Ntrk. 
This behavior is different from that observed in the inclusive \pT\ 
  interval, which may, in turn, reflect the convergence of
  the $v_2(\pT)$ between the \pp\ and \pPb\ data shown
  in the top, right panel of Figure\,\ref{fig::compare_systems}.

\begin{figure}
\centering
\includegraphics[width=0.65\linewidth]{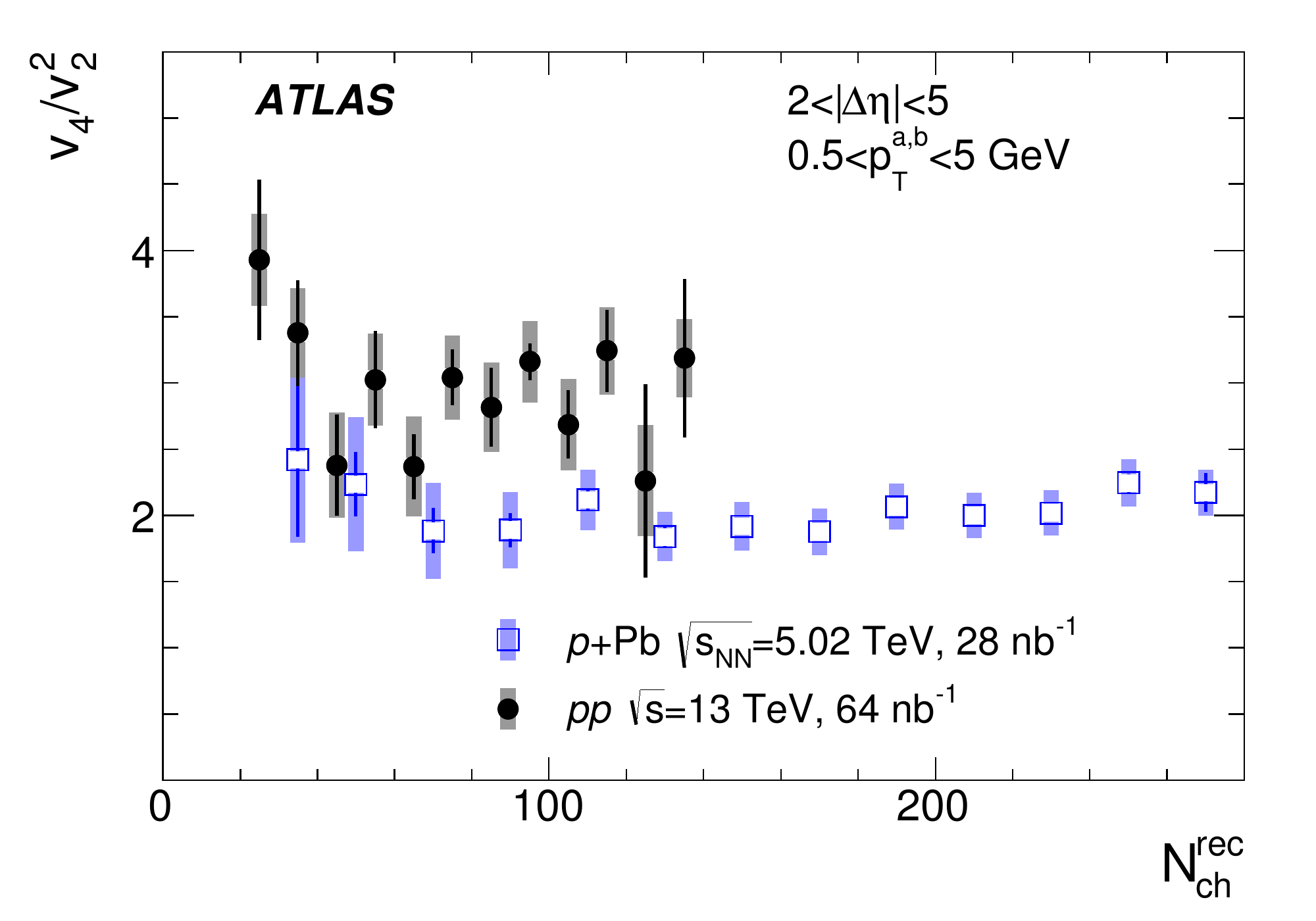}
\caption{
Ratio of \vfour to $\vtwo^{2}$ as a function of \nchrec in the
  13~\TeV\ \pp  and 5.02~\TeV\ \pPb data.
The results are for 0.5<\ptab<5~\GeV. 
The error bars and shaded bands indicate statistical and systematic uncertainties, respectively.
\label{fig:v2_v4_scaling}
}
\end{figure}

Measurements \cite{HION-2012-03,HION-2014-03} and theoretical analyses \cite{Qiu:2012uy,Teaney:2012gu,Teaney:2013dta,Yan:2015jma,Qian:2016fpi}
of the correlations between the Fourier coefficients and
event-plane angles of different flow harmonics in \PbPb\ collisions
have indicated significant ``non-linearity'' resulting from collective
expansion such that the response of the medium to an initial elliptic
eccentricity can contribute to $\cos{(4\phi)}$ modulation of the
produced particles. In \PbPb\ collisions, the non-linear contribution
to $v_4$ is found to dominate over the geometric contribution except for the
most central collisions where the initial-state fluctuations have the
greatest impact. The non-linear contribution to $v_4$ is expected to be
proportional to $v_2^2$ so a comparison of the measured $v_4$
to $v_2^2$ in \pp\ and \pPb\ collisions may be of interest. The
results are presented in Figure\,\ref{fig:v2_v4_scaling}, which shows
$v_4/v_2^2$ versus \Ntrk\ for the 13~\TeV\ \pp\ and the \pPb\
data. 
In the ratio, the correlated systematic uncertainties between 
 the measured $v_4$ and $v_2^2$ cancel.
The ratio is observed to be constant as a function of \Ntrk\ for
both data sets even though the \pPb\ $v_2$ and $v_4$ increase
with \Ntrk. The $v_4/v_2^2$ ratio is observed to be 50\% larger in
the \pp\ data than in the \pPb\ data. Naively, this would indicate a
larger non-linear contribution to $v_4$ in \pp\ collisions
compared to \pPb\ collisions.

\FloatBarrier
\section{Conclusion}
\label{sec:conclusion}
In summary, this \papertype\ presents results of two-charged-particle
  correlation measurements made by ATLAS in $\sqs = 13$ and
  5.02~\TeV\ \pp\ collisions and in 5.02~\TeV\ \pPb\ collisions at the LHC. 
This measurement uses integrated luminosities of \linta for the 
  $\sqs{=}13$~\TeV\ \pp data, \lintb for the $\sqs{=}5.02$~\TeV\ \pp\ data and \lintc 
  for the $\snn{=}5.02$~\TeV\ \pPb\ data.
The 13~\TeV\ measurements represent an extension of results
  presented in Ref.\,\cite{HION-2015-09} using a larger data sample. 
The \pPb\ results are obtained from a reanalysis of Run\,1 data presented 
  in Ref.\,\cite{HION-2013-04} using a template fitting procedure developed for 
  \pp\ collisions and applied in Ref.\,\cite{HION-2015-09}. 
The correlation functions are measured for different intervals of measured 
  charged-particle multiplicity and FCal transverse energy and for different 
  intervals of charged-particle transverse momentum; many of the results 
  are presented for an ``inclusive'' \pT\ interval $0.5 < \pT < 5$~\GeV. 

One-dimensional distributions of per-trigger-particle yields as a
  function of azimuthal angle separation, \Yphi, are obtained from the
  long-range ($|\deta|>2$) component of the correlation functions. 
A template fitting procedure is applied to the \Yphi\ distributions to
  remove the contributions from hard-scattering processes and to measure
  the relative amplitudes $v_{n,n}$ of the sinusoidal modulation of the
  soft underlying event. 
Results for $v_{2,2}$, $v_{3,3}$, and $v_{4,4}$ are obtained for all three 
  colliding systems. An analysis of the factorizability of the $v_{n,n}$ 
  shows good factorization for most of the measured \Ntrk and \pT\ 
  intervals although factorization is observed to break down for the most extreme 
  combinations of \pta\ and \ptb\ in the lowest and highest multiplicity or transverse
  energy intervals. 
Since the $v_{n,n}$ results are observed to be consistent with the presence of 
  single-particle modulation of the per-event $dN/d\phi$ distributions, 
  single-particle $v_n$ values are extracted and plotted versus \Ntrk\ and \pT. 

Comparisons of the $v_2, v_3$ and $v_4$ values between 13 and
  5.02~\TeV\ \pp collisions show no significant variation in these quantities with
  center-of-mass energy. 
As observed in Ref.\,\cite{HION-2015-09}, the $v_2$ values obtained in 
  \pp\ collisions at both energies are observed to be 
  independent of \Ntrk\ within uncertainties for the inclusive \pT\ interval. 
However, for the $1{<}\pT{<}5$~\GeV\ interval a ${\sim}15\%$ decrease in 
  $v_2$ is seen in the lowest \Ntrk\ intervals. 
The \pPb\ $v_2$ values are larger than the  \pp\ $v_2$ values for all 
  multiplicities and are observed to increase slowly with \Ntrk. 
However, the \pPb\ trend appears to converge with the \pp\ values for 
  the lowest multiplicities, at least in the inclusive \pT\ interval. 
For the $1{<}\pT{<}5$~\GeV\ interval, the $v_2(\pT)$ trends do not show the 
  same convergence between \pp\ and \pPb\ results. 
Similar to the results for $v_2$, the \pp\ $v_3$ and $v_4$ values are 
  consistent with being independent of \Ntrk\ within uncertainties and 
  the \pPb\ values are observed to increase with \Ntrk. 
The \pp\ and \pPb\ $v_3$ and $v_4$ values are consistent within 
  uncertainties in the lowest measured \Ntrk\ intervals. 

The \pT\ dependence of the \pp\ and \pPb\ $v_2$ values is similar:
  both rise approximately linearly with \pT\ and reach a maximum near
  3~\GeV. 
The maximum \pPb\ $v_2$ value is approximately 50\% larger than the 
  maximum $v_2$ values for the 13 and 5.02~\TeV\ \pp\ data, which are 
  consistent within uncertainties.   
The \pPb\ $v_3$ and $v_4$ values also increase more rapidly with 
  increasing \pT\ than the corresponding \pp\ values for 
  $\pT < 2$~\GeV, but the \pPb\ $v_3$ values saturate above 3~\GeV\ 
  while the measured 13~\TeV\ \pp $v_3$ values continue to increase with 
  increasing \pT\ over the full range of the measurement. 
A test of the similarity of the \pT\ dependence of the \pp\ and \pPb\ 
  $v_2$ values rescaling \pp\ $v_2$ values shows that the \pp\ and 
  \pPb\ $v_2(\pta)$ distributions are remarkably similar in shape for 
  $\pta{<}5$~\GeV. 

An evaluation of the $v_4/v_2^2$ ratio in the inclusive \pT\ interval 
  shows results that are \Ntrk-independent for both the 13~\TeV\ \pp\ data
  and the \pPb\ data. 
This ratio is observed to be 50\% larger for the \pp\ data than for the \pPb\ data.

The similarities between the \pp\ and \pPb\ results presented here
  suggest a common physical origin for the azimuthal anisotropies. 
The difference in the observed multiplicity dependence of the Fourier
  coefficients likely arises from the different geometry of the \pp\
  and \pPb\ collisions.

\section*{Acknowledgements}

We thank CERN for the very successful operation of the LHC, as well as the
support staff from our institutions without whom ATLAS could not be
operated efficiently.

We acknowledge the support of ANPCyT, Argentina; YerPhI, Armenia; ARC, Australia; BMWFW and FWF, Austria; ANAS, Azerbaijan; SSTC, Belarus; CNPq and FAPESP, Brazil; NSERC, NRC and CFI, Canada; CERN; CONICYT, Chile; CAS, MOST and NSFC, China; COLCIENCIAS, Colombia; MSMT CR, MPO CR and VSC CR, Czech Republic; DNRF and DNSRC, Denmark; IN2P3-CNRS, CEA-DSM/IRFU, France; SRNSF, Georgia; BMBF, HGF, and MPG, Germany; GSRT, Greece; RGC, Hong Kong SAR, China; ISF, I-CORE and Benoziyo Center, Israel; INFN, Italy; MEXT and JSPS, Japan; CNRST, Morocco; NWO, Netherlands; RCN, Norway; MNiSW and NCN, Poland; FCT, Portugal; MNE/IFA, Romania; MES of Russia and NRC KI, Russian Federation; JINR; MESTD, Serbia; MSSR, Slovakia; ARRS and MIZ\v{S}, Slovenia; DST/NRF, South Africa; MINECO, Spain; SRC and Wallenberg Foundation, Sweden; SERI, SNSF and Cantons of Bern and Geneva, Switzerland; MOST, Taiwan; TAEK, Turkey; STFC, United Kingdom; DOE and NSF, United States of America. In addition, individual groups and members have received support from BCKDF, the Canada Council, CANARIE, CRC, Compute Canada, FQRNT, and the Ontario Innovation Trust, Canada; EPLANET, ERC, ERDF, FP7, Horizon 2020 and Marie Sk{\l}odowska-Curie Actions, European Union; Investissements d'Avenir Labex and Idex, ANR, R{\'e}gion Auvergne and Fondation Partager le Savoir, France; DFG and AvH Foundation, Germany; Herakleitos, Thales and Aristeia programmes co-financed by EU-ESF and the Greek NSRF; BSF, GIF and Minerva, Israel; BRF, Norway; CERCA Programme Generalitat de Catalunya, Generalitat Valenciana, Spain; the Royal Society and Leverhulme Trust, United Kingdom.

The crucial computing support from all WLCG partners is acknowledged gratefully, in particular from CERN, the ATLAS Tier-1 facilities at TRIUMF (Canada), NDGF (Denmark, Norway, Sweden), CC-IN2P3 (France), KIT/GridKA (Germany), INFN-CNAF (Italy), NL-T1 (Netherlands), PIC (Spain), ASGC (Taiwan), RAL (UK) and BNL (USA), the Tier-2 facilities worldwide and large non-WLCG resource providers. Major contributors of computing resources are listed in Ref.~\cite{ATL-GEN-PUB-2016-002}.

\printbibliography
\newpage
\begin{flushleft}
{\Large The ATLAS Collaboration}

\bigskip

M.~Aaboud$^\textrm{\scriptsize 137d}$,
G.~Aad$^\textrm{\scriptsize 88}$,
B.~Abbott$^\textrm{\scriptsize 115}$,
J.~Abdallah$^\textrm{\scriptsize 8}$,
O.~Abdinov$^\textrm{\scriptsize 12}$,
B.~Abeloos$^\textrm{\scriptsize 119}$,
O.S.~AbouZeid$^\textrm{\scriptsize 139}$,
N.L.~Abraham$^\textrm{\scriptsize 151}$,
H.~Abramowicz$^\textrm{\scriptsize 155}$,
H.~Abreu$^\textrm{\scriptsize 154}$
% ...
\bigskip
\\
% ...
$^{8}$ Department of Physics, The University of Texas at Arlington, Arlington TX, United States of America\\
% ...
$^{12}$ Institute of Physics, Azerbaijan Academy of Sciences, Baku, Azerbaijan\\
% ...
$^{88}$ CPPM, Aix-Marseille Universit{\'e} and CNRS/IN2P3, Marseille, France\\
% ...
$^{115}$ Homer L. Dodge Department of Physics and Astronomy, University of Oklahoma, Norman OK, United States of America\\
% ...
$^{119}$ LAL, Univ. Paris-Sud, CNRS/IN2P3, Universit{\'e} Paris-Saclay, Orsay, France\\
% ...
$^{137}$ $^{(a)}$ Facult{\'e} des Sciences Ain Chock, R{\'e}seau Universitaire de Physique des Hautes Energies - Universit{\'e} Hassan II, Casablanca; $^{(b)}$ Centre National de l'Energie des Sciences Techniques Nucleaires, Rabat; $^{(c)}$ Facult{\'e} des Sciences Semlalia, Universit{\'e} Cadi Ayyad, LPHEA-Marrakech; $^{(d)}$ Facult{\'e} des Sciences, Universit{\'e} Mohamed Premier and LPTPM, Oujda; $^{(e)}$ Facult{\'e} des sciences, Universit{\'e} Mohammed V, Rabat, Morocco\\
% ...
$^{139}$ Santa Cruz Institute for Particle Physics, University of California Santa Cruz, Santa Cruz CA, United States of America\\
% ...
$^{151}$ Department of Physics and Astronomy, University of Sussex, Brighton, United Kingdom\\
% ...
$^{154}$ Department of Physics, Technion: Israel Institute of Technology, Haifa, Israel\\
$^{155}$ Raymond and Beverly Sackler School of Physics and Astronomy, Tel Aviv University, Tel Aviv, Israel\\
% ...
$^{a}$ Also at Department of Physics, King's College London, London, United Kingdom\\
$^{b}$ Also at Institute of Physics, Azerbaijan Academy of Sciences, Baku, Azerbaijan\\
$^{c}$ Also at Novosibirsk State University, Novosibirsk, Russia\\
$^{d}$ Also at TRIUMF, Vancouver BC, Canada\\
$^{e}$ Also at Department of Physics {\&} Astronomy, University of Louisville, Louisville, KY, United States of America\\
% ...
\end{flushleft}


\end{document}